\documentclass[]{aastex631}

\usepackage{rotating}

\accepted{November 11, 2024}

\begin{document}

\title{Rising Near-Ultraviolet Spectra in Stellar Megaflares}

\author[0000-0001-7458-1176]{Adam F. Kowalski}
\affiliation{Department of Astrophysical and Planetary Sciences, University of Colorado Boulder, CO 80305, USA}
\affiliation{National Solar Observatory, Boulder, CO 80303, USA}
\affiliation{Laboratory for Atmospheric and Space Physics, Boulder, CO 80303, USA}

\author[0000-0001-5643-8421]{Rachel A. Osten}
\affiliation{Space Telescope Science Institute, Baltimore, MD 21218, USA}

\author[0000-0002-0412-0849]{Yuta Notsu}
\affiliation{Laboratory for Atmospheric and Space Physics, Boulder, CO 80303, USA}
\affiliation{National Solar Observatory, Boulder, CO 80303, USA}

\author[0000-0001-5974-4758]{Isaiah I. Tristan}
\affiliation{Department of Astrophysical and Planetary Sciences, University of Colorado Boulder, CO 80305, USA}
\affiliation{Laboratory for Atmospheric and Space Physics, Boulder, CO 80303, USA}
\affiliation{National Solar Observatory, Boulder, CO 80303, USA}

\author[0000-0002-2240-2452]{Antigona Segura}
\affiliation{Instituto de Ciencias Nucleares, Universidad Nacional Autónoma de México, Circuito Exterior s/n, Ciudad Universitaria, Ciudad de México, México}

\author[0000-0003-0332-0811]{Hiroyuki Maehara}
\affiliation{Subaru Telescope Okayama Branch Office, National Astronomical Observatory of Japan, National
Institutes of Natural Sciences, 3037-5 Honjo, Kamogata, Asakuchi, Okayama 719-0232, Japan}

\author[0000-0002-1297-9485]{Kosuke Namekata}
\affiliation{The Hakubi Center for Advanced Research, Kyoto University, Kyoto 606-8302, Japan}
\affiliation{Department of Physics, Kyoto University, Kitashirakawa-Oiwake-cho, Sakyo-ku, Kyoto, 606-8502, Japan}
\affiliation{Division of Science, National Astronomical Observatory of Japan, NINS, Osawa, Mitaka, Tokyo, 181-8588, Japan}

\author[0000-0003-3085-304X]{Shun Inoue}
\affiliation{Department of Physics, Kyoto University, Kitashirakawa-Oiwake-cho, Sakyo-ku, Kyoto, 606-8502, Japan}

\begin{abstract}
Flares from M-dwarf stars can attain energies up to $10^4$ times larger than solar flares but are generally thought to result from similar processes of magnetic energy release and particle acceleration.
Larger heating rates in the low atmosphere are needed to reproduce the shape and strength of the observed continua in stellar flares, which are often simplified to a blackbody model from the optical to the far-ultraviolet (FUV). The near-ultraviolet (NUV) has been woefully undersampled in spectral observations despite this being where the blackbody radiation should peak.  We present \emph{Hubble Space Telescope} NUV spectra in the impulsive phase of a flare with $E_{\rm{TESS}} \approx 7.5 \times 10^{33}$ erg and a flare with $E_{\rm{TESS}} \approx 10^{35}$ erg and the largest NUV flare luminosity observed to date from an M star.  The composite NUV spectra are not well represented by a single blackbody that is commonly assumed in the literature.  Rather, continuum flux rises toward shorter wavelengths into the FUV, and we calculate that an optical $T=10^4$ K blackbody underestimates the short wavelength NUV flux by a factor of $\approx 6$.  We show that rising NUV continuum spectra can be reproduced by collisionally heating the lower atmosphere with beams of $E \gtrsim 10$ MeV protons or $E \gtrsim 500$ keV electrons and flux densities of $10^{13}$ erg cm$^{-2}$ s$^{-1}$.  These are much larger than canonical values describing accelerated particles in solar flares.
\end{abstract}


\section{Introduction} \label{sec:intro}

Stellar flares are the most dramatic examples of variability that a cool star experiences while on the main sequence. 
Current understanding \citep{Kowalski2024LRSP} is that stellar flares are produced as a result of magnetic reconnection in the tenuous outer atmosphere of the star, which happens as a result of the jostling and colliding of magnetic fields that are rooted in the lower stellar atmosphere. 
Flares involve all layers of the stellar atmosphere, 
with a variety of physical processes, from plasma heating to particle acceleration to mass motions \citep{BenzGudel2010, Notsu2024}. 
These three factors result in flare emissions being produced across the electromagnetic spectrum. 
While flares on the Sun are the best studied due to 
the Sun's  proximity, flaring itself is a product of magnetic activity and can be seen on a variety of stars with outer convective envelopes, from pre-main sequence to young solar-type to evolved
stars \citep{Yang2019, Okamoto2021}. 
Apart from solar studies, M dwarfs are the most studied for stellar flares, due to a combination
of population statistics \citep[M dwarfs are the most common type of star in the nearby solar neighborhood,][]{Bochanski2010} and flare statistics 
\citep[M dwarfs tend to have high flaring rates;][]{Candel2014,Yang2019}.

Flare events can have prodigious releases of energy of up to 10$^{36}$ erg, or more than 10,000 times bigger than the largest solar flares \citep[e.g.,][]{Osten2016,Maehara2012}.
In particular, the flare radiation in the ultraviolet (UV) can have significant impacts on nearby potentially habitable planets, which may be either helpful \citep[e.g.][]{Rimmer2018} or harmful \citep[e.g.][]{Segura2010,Tilley2019}. 
The UV spectral region is key to understanding
the impacts that flares have on exoplanet habitability by examining 
their effects on the atmosphere under a variety of assumptions about atmosphere composition and characteristics \citep{Segura2018}.

In early studies of M dwarf flares, \citet{HF92} established that the blue-optical flare spectral energy distribution (SED) was consistent with a blackbody of temperature around 10$^{4}$ K. 
This model was based largely on FUV and optical spectra in the impulsive phase of a flare first studied in \citet{HP91}, and it has remained lore in the stellar flare community. 
The original explanation of the broadband increase, of reprocessing of upper atmospheric (coronal) radiation, was demonstrated in subsequent papers \citep{Allred2006} to be negligible compared to the energy deposited in the lower atmosphere by electron beams.
More recent high-cadence, low-resolution flare spectral atlases have shown the predominance in the blue-optical for a continuum component consistent with such a blackbody, as well as hydrogen Balmer jump emission and higher order Hydrogen Balmer lines \citep{Kowalski2010, Kowalski2013}. 
Despite the evidence for multiple components, the functional form of a single-temperature blackbody has persisted in explaining flare continuum emission from the UV to optical wavelengths \citep[e.g.][among many others]{Loyd2018, Gunther2020, Howard2020, Okamoto2021,Berger2024, Feinstein2024}.  Spectroscopic  \citep{Froning2019, Howard2023} and broadband \citep{Brasseur2023, Paudel2024} investigations have started to consider multi-component models from the FUV to the NIR, but spectra that span the Balmer jump can be helpful in limiting the parameter space and breaking degeneracies \citep{Kowalski2019HST}.

The optical and ultraviolet portions of the flare radiation trace emission produced in the lower stellar atmosphere. It was perplexing given the generally good agreement between solar and stellar flares in terms of scaling relations for  plasma heating 
\citep[e.g.][]{Shibata1999, Shibata2002}  
that the large optical continuum enhancements seen in M dwarf flares could not be reproduced using 
solar flare models \citep{Allred2006}.
Current state-of-the-art radiative-hydrodynamic (RHD) models of solar and stellar flares \citep{Allred2005,Kowalski2017Broadening} have demonstrated an ability to reproduce continuum flare features as well as the strongest line emission, in the NUV through blue-optical 
\citep[see also][]{Brasseur2023} through the action of accelerated particles impacting the lower stellar atmosphere. 
Solar studies generally utilize hard X-ray observations of nonthermal bremsstrahlung emission and radio observations to 
diagnose the presence and action of accelerated particles \citep{Holman2011, Kontar2011, White2011}.
The former is not available for stellar studies due to sensitivity issues, and radio observations in the microwave
regime have not had the frequency coverage to constrain the properties of the accelerated particles in stellar flares \citep{Osten2005}. 
Earlier studies \citep{Gudel1996}
have suggested a disconnect in particle acceleration relative to plasma heating for the stellar flares studied versus solar flares.  Also, the large gap between  solar and stellar optical continuum flares in energy-duration diagrams can be attributed to physical scaling relations that imply larger magnetic fields (and flaring volumes) in active stars \citep[][and see also studies of X-ray superflares, such as \citealt{Favata2000} and \citealt{Osten2016}]{Maehara2015, Namekata2017}.
Confirmation that the UV spectral region can diagnose the characteristics of magnetic energy and accelerated particles in stellar flares
would open new regimes for exploring solar-stellar connections.

Stellar flares have been studied at UV wavelengths predominantly in the FUV \citep[e.g.,][]{Hawley2003,France2016,Loyd2018} for many decades, and current generations of planet transit hunting telescopes like TESS and Kepler routinely pick up the optical counterparts of these events \citep{Davenport2016,Gunther2020}.  
Despite this seemingly advanced topic, recent discoveries have thrown doubt on how well we understand the stellar flare process.  
Some of this arises from the relative lack of observations in the NUV wavelength region, despite this being where such a 10$^{4}$ K blackbody would peak. 
Surprisingly, in a recent multi-wavelength flare study, only 25\% of NUV flares on M dwarfs had
an optical counterpart \citep{Paudel2024}, and some NUV flares clearly show a much greater response than a $T\approx9000-10^4$ K blackbody would predict from the available optical constraints \citep[][and see also \citealt{Kowalski2019HST}]{Jackman22, Brasseur2023, Jackman2023}.
These results motivate renewed scrutiny of 
the NUV spectral region in stellar flares
and form the basis of an 
HST Treasury program designed to use the NUV as a ``fulcrum'' to bridge 
the more often-studied FUV and the very well-studied optical regions
(HST Guest Observer 17464 ``From High-Energy Particle Beam Heating in Stars to Ozone Destruction in Planets: NUV Spectra as the Fulcrum for a Comprehensive Understanding of Flaring M Dwarf Systems''). 
In the following sections, we describe early
results from this Treasury program, highlighting two remarkable events and what they can tell us about physical processes at work in stellar flares.
Section~\ref{sec:data} describes the data, \S~\ref{sec:analysis} the light curve and spectral analysis, 
\S~\ref{sec:models} application of RHD models to understand the continuum and line emission,
\S~\ref{sec:discussion}
discusses the results,
and \S~\ref{sec:conclusions} concludes. Appendix \ref{sec:data_appendix} describes the data reduction and Appendices \ref{sec:supplementaryLCs} and \ref{sec:supplementary_spectra} present light curves and flare spectra that supplement the figures in the main text.

\section{Data} \label{sec:data}

CR Dra is a low-mass, M-dwarf star that is well known to flare in the NUV \citep{Welsh2006, Million2016, Jackman2024}.  It is a spectroscopic binary with a trigonometric parallax distance of $20.15 \pm 0.17$ pc \citep{GaiaDR2, GaiaDR3}, consisting of a M0 star and a lower mass companion \citep{Tamazian2008}.  The combined spectral type is M1e \citep{PMSU1, PMSU2}, and there is a single source in Gaia with a $G$-band absolute magnitude that is well above the main sequence at the color of other M1 stars \citep{Kowalski2024LRSP}\footnote{The fundamental properties of this system, such as the orbital period and even its distance, are not yet agreed upon \citep{Tamazian2008, Shkolnik2014, Sperauskas2019}.}.  

The Cosmic Origins Spectrograph (COS) on the \emph{Hubble Space Telescope} (HST) observed two extraordinary (``mega''\footnote{Merriam-Webster defines ``mega'' as ``great; large'' or ``greatly surpassing others of its kind''. Based on the properties of the two events outlined in this paper, these flares surely have earned the ``megaflare'' designation.}) flares on CR Dra during the observations for Guest Observer (GO) Treasury Program 17464.  The NUV G230L grating was employed with the 2950 \AA\ central wavelength, giving spectral coverage at $\lambda \approx 1683 - 2082$ \AA\ in Stripe A (NUVA) and $\lambda \approx 2773 - 3170$ \AA\ in Stripe B (NUVB).  The linear dispersion is $0.39$ \AA\ pixel$^{-1}$, and the resolving power, $R$, varies from 2100 to 3900.  The photons are time-tagged which allows for spectra to be extracted over any time interval.  The Transiting Exoplanet Survey Telescope \citep[TESS;][]{Ricker2015} simultaneously observed CR Dra  at 20~s and 120~s cadences in Sectors 76 and 78 in Cycle 6.  Appendix \ref{sec:data_appendix} describes the data processing in detail.

\section{Light Curve and Spectral Analysis} \label{sec:analysis}

The broadband HST light curves of the two megaflare events are shown in Figure \ref{fig:lcs}.  
These light curves exclude the Mg II $h$ and $k$ emission lines and the pixels within each stripe that are affected by detector shadow (Section \ref{sec:hstdata}).  The light curves for NUVA and NUVB are given as average flare-only flux densities over the respective wavelength ranges, $\big<f^{\prime}_{\lambda}\big>$, where the prime symbol indicates a preflare flux is subtracted.  
Flare Event 1 consists of many extraordinary variations superimposed on two major peaks.  The HST observations stop during the flare due to the end of the orbit.  This time corresponds to the end of the fast decay phase in the $\Delta t = 20$~s TESS light curve (Appendix \ref{sec:supplementaryLCs}). The energy in the TESS bandpass is $E_{\rm{TESS}} = 7.5 \times 10^{33}$ erg, and the peak luminosity is $6.1 \times 10^{30}$ erg s$^{-1}$, making this an extremely energetic event.  Flare Event 2 is over 10x more energetic having $E_{\rm{TESS}} \approx 10^{35}$ erg.  Flare Event 2 is the most luminous NUV flare that has been reported (to our knowledge) on an M-dwarf star; the spectral luminosity in NUVA is over a factor of 10 larger than the most luminous flare spectra compiled from the  International Ultraviolet Explorer (IUE) / Short-Wavelength Prime (SWP) Spectrograph in \citet{Phillips1992}.  The rise phase increases the NUVB stellar flux ($\approx 1.24\times10^{-14}$ erg cm$^{-2}$ s$^{-1}$ \AA$^{-1}$) by a factor of $\gtrsim 200$, but the buffer filled as HST stopped observations due to an overlight condition.  After this time, the broadband optical/NIR response continues to increase, and the TESS luminosity reaches $8.8 \times 10^{31}$ erg s$^{-1}$ at peak  (Appendix \ref{sec:supplementaryLCs}). 
 Flare Event 2 is about 25 times more impulsive than Flare Event 1 in the TESS band, where we follow \citet{Kowalski2013} and define the impulsiveness index of a light curve as the flare-only peak flux divided by the full-width-at-half maximum (FWHM) duration, $t_{1/2}$.

Within each flare event, the NUVA light curves are more impulsive and show peak spectral flux densities that are larger than at the longer wavelengths of NUVB  (Figure \ref{fig:lcs}).  The impulsiveness index is $2-2.3$x larger and peak fluxes are  $\approx 1.4-1.5$x higher in NUVA during the two major peaks in Flare Event 1.  By the end of the HST observations of Flare Event 2, the peak flux in NUVA is a factor of $\approx 1.7$ larger than NUVB.  These properties are qualitatively similar to broadband GALEX/FUV (1350--1750 \AA) and GALEX/NUV (1750--2800\AA) filter ratios during stellar flares \citep{Robinson2005, Welsh2006, Million2016, Fleming2022, Berger2024}.  The impulsiveness differences between NUVA and NUVB are also qualitatively similar to  \citet{Hawley2003}, who found a power-law relationship between the time-evolution of FUV continuum regions (averaged over $\lambda = 1420-1452$ \AA\ and $1675-1710$ \AA) and the time-evolution of the $U$ band  ($\lambda \approx 3200-4000$ \AA) flux.  Over the entire impulsive phase light curve of Flare Event 1, we calculate the power-law index, $\alpha$, from $\log_{10} \big< f^{\prime}_{\lambda,\rm{NUVA}} \big> = \beta + \alpha \log_{10} \big< f^{\prime}_{\lambda,\rm{NUVB}} \big>$ as $\alpha \approx 1.5$.  Over the rise phase of Flare Event 2, we calculate $\alpha \approx 1.3$.  These values are similarly $>1$, like the value of $\alpha = 1.65$ that is reported in the \citet{Hawley2003} data.  A nonlinear relationship means that the flux ratios vary as functions of the fluxes, which has remained unexplained in $U$-band and FUV comparisons.  However, previous interpretations of broadband UV flare data \citep[e.g.,][]{Hawley2003, Mitra2005, Robinson2005, Welsh2006, Berger2024} have been severely limited to simple models, such as emission line slabs or single-temperature blackbody functions, due to the lack of an NUV spectral fulcrum connecting the blue edge of the $U$ band to the FUV.

For the first time, we present ultraviolet spectral observations at $\lambda \lesssim 3200$ \AA\ that show there are two blackbody color temperatures in the NUV with $T_{\rm{NUVA}}$ $>$ $T_{\rm{NUVB}}$.  The spectra and the fitting results from the second major peak in Flare Event 1 and from the rise phase of Flare Event 2 are shown in Figure \ref{fig:spectra}.   Continuum radiation dominates at both $\lambda \lesssim 3200$ \AA\ (in our NUVB stripe) and at $\lesssim 2100$ \AA\ (in our NUVA Stripe, which some may consider as part of the FUV or mid-ultraviolet) in the impulsive phases of these flares.   The inferred NUVB continua are strikingly flat within this range: we use the non-linear least squares Levenberg-Marquardt algorithm and fit blackbody temperatures of $\hat{T} \approx 9000-10,000$ K.  The NUVA spectra are overall brighter than the NUVB spectra, as noted in the continuum light curves (Figure \ref{fig:lcs}), and they exhibit an ascent towards shorter wavelengths.  We independently fit blackbody temperatures to NUVA and find that $\hat{T}_{\rm{NUVA}}$ ranges from $\approx 16,000$ K to nearly $18,000$ K;  see Appendix \ref{sec:supplementary_spectra} for several other flare spectra and the corresponding fits.  A two-temperature blackbody fit to NUVA and NUVB results in a high-temperature component that is only a few hundred degrees hotter than the fit to only NUVA.
A single-temperature fit to the ratio, $\big< f^{\prime}_{\lambda, \rm{NUVA}}\big>/\big<f^{\prime}_{\lambda, \rm{NUVB}} \big>$, of the wavelength-averaged flux densities in the spectrum of Flare Event 2 gives a blackbody color temperature of $\hat{T}_{\rm{NUV}} = 14,790 \pm 430$ K.  This is  not representative of the continuum shape within either NUV stripe, thus demonstrating that one filter ratio spanning a large wavelength range across the UV does not accurately characterize the flare SED.

There are probably many faint emission lines  \citep{Cook1979,Doyle1992} that blend together and form a pseudo-continuum on top of the bona-fide continuum radiation.  However, the contribution appears to be energetically small in the spectra of these flares: We calculate that the continuum fits comprise $97-99$\% of the wavelength-integrated fluxes within NUVA and NUVB.
  This is expected based on knowledge from an echelle flare spectrum of the impulsive phase of an M-dwarf flare at $\lambda = 3300 - 3800$ \AA, which shows a very small contribution from emission lines in comparison to a $T \approx 11,000$ K blackbody continuum model fit \citep{Fuhrmeister2008}.  We have identified regions in the HST spectra (indicated by the gray bands in Figure \ref{fig:spectra}) that correspond to the low-points between obvious emission lines \citep[e.g.,][]{Doyle1992}. These regions best represent the bona-fide continuum fluxes, which are used in the fitting procedure.  Notably, the best-fit blackbody temperatures are largely robust to the choice of fitting windows outside of the emission lines of Al III (1854.716 \AA, 1862.790 \AA) in NUVA (indicated by vertical dashed lines) and Mg II $h$ and $k$ in NUVB.  The other two vertical dashed lines in NUVA indicate Si II $\lambda$1808.0 and $\lambda$1816.9, which are among the brightest in the pre-flare and in gradual phase spectra of solar flares \citep{Doyle1992,Simoes2019}.  

The gradual temporal evolution of the Mg II emission line fluxes further justifies attributing the dominant source of radiative energy to continuum radiation.  We integrate over the wavelengths of the resonance lines and the nearby subordinate lines, which also brighten in the flares, and we show the continuum-subtracted flux evolution at the 20~s cadence of TESS in Figure \ref{fig:lcs}.   In Flare Event 1, there is a striking lack of similarity between the Mg II emission line response and the two main wavelength-integrated peak fluxes, which are dominated by the continuum radiation.  The Mg II light curve rises and peaks more gradually, and it remains significantly elevated above preflare levels while still decaying at the start of the next HST orbit (Appendix \ref{sec:supplementaryLCs}).

\section{Radiative-hydrodynamic Flare Modeling} \label{sec:models}

Previous RHD modeling \citep{Kowalski2022Frontiers} of the multi-wavelength observations of the impulsive phase of the Great Flare of AD Leo \citep{HP91, HF92} provides a starting point for explaining the new HST spectra in terms of collisional heating from nonthermal, power-law electron beams.  
\citet{Kowalski2022Frontiers} find that several combinations of models with large electron beam energy flux densities are possible explanations for the hydrogen Balmer spectra and broad-wavelength constraints, including the rather flat shape of the FUV continuum, in the Great Flare.  We use the publicly available output\footnote{10.5281/zenodo.10929514} from a grid \citep{Kowalski2024} of RADYN \citep{Carlsson1992B, Carlsson1995, Carlsson1997, Allred2015, Carlsson2023} stellar flare models  and perform a complete parameter space search for radiative surface flux continuum spectra that are able to explain the NUVA and NUVB shapes in Flare Event 2 (Figure \ref{fig:spectra}, bottom). The shapes of the model spectra are determined by the ratios of the detailed continuum fluxes at $\lambda = 1800$ \AA\ and $2069$ \AA\ in NUVA and at $\lambda = 2830$ \AA\ and $3080$ \AA\ in NUVB.  These are compared to the ratios of  flare-only fluxes in the data at averages over $\lambda = 1778.7 -  1805$ \AA\ and $2065.16- 2080.14$ \AA\ in NUVA, and over $\lambda = 2820 - 2850$ \AA\ and $3011 - 3170$ \AA\ in  NUVB.

Only the most energetic electron beam heating model (\texttt{cF13-500-3}) with an energy flux density of $10^{13}$ erg cm$^{-2}$ s$^{-1}$, a low-energy cutoff of $E_c = 500$ keV, and a number-flux power-law index of $\delta =3$ above this low-energy cutoff can account for the rising slope of Flare Event 2 within NUVA.  This continuum model is shown in Figure \ref{fig:RHD}.  However, the model fails to fully account for the flux at the longer wavelengths in NUVB.   We thus follow \citet{Kowalski2022Frontiers} and fit linear superpositions of every combination of two RHD model spectra to the four continuum regions in the data, 

\begin{equation}
    f^{\prime}_{\lambda, \rm{obs}} = \frac{R^2_{\rm{star}}}{d^2} \big( \hat{X}_1 F^{\prime}_{\lambda, 1} + \hat{X}_2 F^{\prime}_{\lambda, 2} \big)
\end{equation}

\noindent where $\hat{X}_1$ is the best-fit filling factor of RHD model component 1 and $F^{\prime}_{\lambda,1}$ is the radiative surface flux model of component 1 with the pre-flare subtracted, $R_{\rm{star}} = 4 \times 10^{10}$ cm is an approximate radius of the flare star, and $d$ is the distance to CR Dra (see \citet{Kowalski2022Frontiers} for other details).  
The fitting is performed to 98\% of the observed flux densities, which allows the models to better account for the continuum distribution that underlies a pseudo-continuum pedestal of faint emission lines within the noise.  
The filling factor of each component is directly proportional to the area of the respective flaring source at the star. We report $\hat{X}_{\rm{rel}} = \hat{X}_2 / \hat{X}_1$, and we leave out the factor $\hat{X}_1\frac{R^2_{\rm{star}}}{d^2}$ from the labels of the model combinations in figure legends.

As in the modeling of the AD Leo Great Flare, a combination of the \texttt{cF13-500-3} with an electron beam flux density of $2 \times 10^{12}$ erg cm$^{-2}$ s$^{-1}$, a smaller low-energy cutoff ($E_c = 37$ keV), and hard power-law index, $\delta = 2.5$, gives an adequate explanation for the overall properties of the full UV continuum range in the new data. For Flare Event 2 (Figure \ref{fig:RHD}), the best-fit filling factors, $\hat{X}$, of each model imply a ratio of the area of the low-flux density model (\texttt{m2F12-37-2.5}) to the area of the high-flux density model (\texttt{cF13-500-3}) of $\hat{X}_{\rm{rel}} = \hat{X}_2 / \hat{X}_1  = \hat{X}_{\rm{low}}/ \hat{X}_{\rm{high}} \approx 3.4$.    The same combination of models results in a value of 
$\hat{X}_{\rm{rel}} = 7.3$ for the spectrum of Flare Event 1 (Appendix \ref{sec:supplementary_spectra}), which suggests that the combination is a rather flexible model for the impulsive phase of very energetic flares.   Section \ref{sec:discussion} discusses interpretations of the two-component RHD model fitting.

Though the two-component RHD model of the NUVB is still not flat enough, detailed radiative transfer modeling of the Mg II emission lines  lends credence to the general approach of superposing these RHD models.  We use the RH code \citep{Uitenbroek2001} to calculate the spectra with the 10-level quintessential model Mg II atom \citep{Leenaarts2013} and the updated quadratic Stark damping \citep{Zhu2019} from the STARK-B database \citep{Sahal1995MgII, Sahal2011}\footnote{https://stark-b.obspm.fr/}.  Figure \ref{fig:MgII} shows representative spectra of Mg II that are calculated from snapshots of the \texttt{cF13-500-3} and the \texttt{m2F12-37-2.5} models.  The model spectra are convolved with the G230L line spread function and are binned to the linear dispersion ($0.39$ \AA\ pix$^{-1}$).  A modest scaling of the total model shows general agreement with the observations of the early rise phase of Flare Event 2.  The Mg II $h$ and $k$ emission lines at high spectral resolution are notoriously difficult to reproduce in solar flare models \citep{Rubio2017, Zhu2019, Dalda2023, Kerr2024A}, but the models here appear to have a rather reasonable amount of radiative energy loss at optically thick lines relative to the radiative energy loss through the continuum \citep[see][for comparisons of the model grid to hydrogen Balmer lines]{Namekata2020, Kowalski2022Frontiers, Kowalski2024}.

\section{Discussion} \label{sec:discussion}

Until now, there has been only a handful of NUV spectra during the impulsive phase of stellar flares \citep{Robinson1995, Hawley2007, Kowalski2019HST, Jackman2024}.  The AD Leo Great Flare spectra in the NUV from IUE/Long-Wavelength Prime (LWP) spectrograph are saturated in the impulsive phase; thus, spectra have been merged from different phases of the flare with heterogeneous exposure times.  This merged flare spectrum, shown in Figure \ref{fig:RHD}, has been widely used as a template flare SED in exoplanet photochemistry modeling \citep{Segura2010, Venot2016, Tilley2019, Ridgway2023} and assessments of UV-dependent pathways to the origin of life \citep{Ranjan2017, Armas2023}. In the fiducial Great Flare template,  the NUV is more luminous than the FUV.  
The emission line fluxes are larger relative to the continuum radiation than in the CR Dra flare spectra, as expected for a spectrum that was obtained from the gradual decay phase \citep{HP91}.   
Based on spectra and photometry in the optical and $U$-band of other flares \citep{HF92, Hawley1995, Hawley2003, Fuhrmeister2008, Kowalski2013}, it was expected that with sufficient signal-to-noise and a large enough flare, the spectral flux density in the impulsive phase would be seen to clearly peak and turn over at  $\lambda \approx 3000$ \AA, like a $T \approx 10^4$ K blackbody or blackbody-like spectrum within the hydrogen Balmer continuum.  

Now, we see that the inferred continuum radiation in the short-wavelength NUV rises well above the continuum fluxes in the longer-wavelength ultraviolet around $\lambda = 3000$ \AA\ in two megaflare events on CR Dra.   We conclude that there is a large amount of energy from (at least some) stellar flares  in the impulsive phase that is radiated in the short-wavelength NUV continuum -- and by a sensible extrapolation, in the FUV as well.  This energy has previously been unaccounted for through the AD Leo Great Flare template (Figure \ref{fig:RHD}), through blackbody extrapolations from optical photometry \citep[e.g.,][]{Shibayama2013, Gunther2020}, and through RHD extrapolations based on Balmer jump strengths \citep{Kowalski2019HST, Kowalski2022Frontiers}. In Figure \ref{fig:summaryfig}, we show a $T=10^4$ K blackbody anchored to the  TESS flare-only flux (blue star symbol) averaged over UTC 2024-05-17 11:45:00.85 to 11:47:00.85 during Flare Event 2.  The HST/COS spectra in this figure correspond to the same time interval, and the observed fluxes within NUVA and NUVB are $6.1$x and $2.6$x larger, respectively,  than the extrapolation of the $T=10^4$ K blackbody.

Hotter blackbody color temperatures, ranging from $T \approx 15,000$ K to $T\approx 40,000 \pm 10,000$ K, have been recently reported in several stellar flare events with spectra at shorter wavelengths in the FUV, either at $\lambda \lesssim 1700$ \AA\ or at $\lambda \lesssim 1400$ \AA\ \citep{Loyd2018Hazmat, Froning2019, Feinstein2022, MacGregor2021}.  We suggest that the increasing values of $f_{\lambda}$ within the NUVA range represent the long-wavelength extension of this phenomenon, which is the most luminous component of the flare radiation in our data.  Even with two-temperature blackbody fits (Section \ref{sec:analysis}), we do not find  temperatures larger than 20,000 K within the rise phase spectra of Flare Event 2.  Thus, there is no evidence in our spectra that is suggestive for extremely large temperatures near $T = 30,000 -50,000$ K, which have been claimed in the optical \citep{Howard2020} and the FUV \citep{Robinson2005,Froning2019} in other flares.  As is well known, the FUV has many bound-free edges and bright resonance lines (C II, Si IV, C IV) from metals that make interpretation difficult, especially in broadband data; the quiescent FUV spectrum of a  M1 star has comparably large blackbody temperatures as in flares \citep{Feinstein2022}.  The NUVA, however, is predominantly affected by H I bound-free opacities at the expected temperatures  of flare chromospheres  \citep[e.g.,][Kowalski 2025, in prep]{Garcia2005}. Thus, the NUVA wavelength range provides straightforward comparisons to currently available model predictions.  The simultaneous spectral constraints within the NUVB range connect the continuum radiation that rises into the FUV to the flare flux at the blue-edge of the $U$-band, which has traditionally been used in stellar flare studies over many decades \citep[e.g.,][]{Lacy1976}.  
In the smaller flares of \citet{Hawley2003} (\emph{cf}. flare 8 in their Fig 10), the FUV continuum fluxes are clearly fainter than the $U$-band fluxes.  However, we note that the FUV is well above their $T\approx 9500$ K blackbody fit;  less energetic and less luminous flares \citep[e.g., similar to those in][]{Kowalski2019HST} will be analyzed from the HST Treasury program in future work.

Two phenomenological blackbody color temperatures are necessary to separately account for the blue-optical ($\lambda = 4000-4800$ \AA) and red-optical ($\lambda = 5900-7000$ \AA) continuum spectra in stellar flares \citep{Kowalski2013, Kowalski2016};  see for example the IF3 event peak spectrum in Fig. 32 of \citet{Kowalski2013} and the accompanying discussion.  The peak flare-only optical spectrum of this flare is shown on Figure \ref{fig:summaryfig}. 
In this event, a cooler blackbody can be fit to the redder wavelengths closer to the head of the Paschen continuum, while blue-optical wavelengths at $\lambda = 4000-4800$ \AA\ rise above this;  a hotter blackbody model temperature  is needed \citep{Kowalski2013}.  We have shown that a similar empirical phenomenon occurs across the NUV beginning at wavelengths just shortward of the $U$-band.  This potential connection between blackbody modeling in different wavelength regimes suggests that the NUV and optical continua in the impulsive phases of stellar flares are similarly affected by a common opacity source.   H I bound-free opacity, as in our RHD models with very large electron beam heating rates \citep{Kowalski2017Broadening}, is dominant  in both regimes, while other opacity sources such as H$^{-}$ are not.  A best-fit combination of the RHD models (Section \ref{sec:analysis}) is shown in Figure \ref{fig:summaryfig} and predicts generally similar properties as the IF3 optical peak spectrum: a small Balmer jump and decreasing color temperatures\footnote{In each model component, the optical continua are also ``mulithermal'' \citep[][Kowalski 2025, in prep]{Kowalski2023}.} (not shown) toward the wavelengths of the TESS bandpass ($\lambda \gtrsim 6000$ \AA).  Although simultaneous optical spectra are not available for Flare Event 2 to constrain the Balmer jump and optical emission lines, the observed TESS flare-only flux in Figure \ref{fig:summaryfig} suggests that the extrapolation of the RHD model to the NIR is relatively robust.   The \texttt{cF13-500-3} component in the two-model fit extends almost precisely to the TESS data constraint, while the two-component fit to the NUV predicts a TESS filter-weighted flux that is about twice too large.  RHD models with smaller surface fluxes than the \texttt{m2F12-37-2.5} and cooler blackbody curves with $T \approx 4200$ K, which better account for the NUVB shape in two-component blackbody fitting (Section \ref{sec:analysis}), predict 
optical fluxes far above the TESS constraint.  Also, much smaller model surface fluxes require filling factors that are larger than the visible stellar atmosphere, which is probably unrealistic.    Detailed investigation into the NUV and TESS relationship in all of the flares in the Treasury program will be the subject of a second paper (Osten et al, in prep).

Other sources of large heating rates that are similar to the $E_c=500$ keV model may result in alternative explanations of the CR Dra megaflare spectra.  We leverage the new and unique capabilities of the \texttt{FP} \citep{Allred2020} and \texttt{FP+RADYN} codes \citep{Allred2022,Kerr2023OZ,Kerr2024A} to calculate the transport of proton beam energy and momentum in an M-dwarf atmosphere.  As generally expected from the equations for energy loss in a cold target \citep[e.g.,][see Fig. 25 of \citealt{Kowalski2024LRSP}]{Emslie1978,HF94}, we calculate that a proton beam with $E_c=10$ MeV, a power-law index of $\delta = 3$, and an energy flux density of $10^{13}$ erg cm$^{-2}$ s$^{-1}$ above the cutoff produces a  thermal response in the low stellar atmosphere that is similar to the fully relativistic, high-energy electron beam heating model (\texttt{cF13-500-3}). Thus, a superposition of a high-energy proton beam simulation with a lower energy electron beam simulation (e.g., the \texttt{m2F12-37-2.5} model) could plausibly serve as an explanation for the megaflare spectra.  It is interesting that the lower atmospheric impact sites of accelerated protons in solar flares are significantly displaced from the spatial locations of hard X-ray footpoint sources, which are associated with nonthermal electron energy deposition \citep{Hurford2006}.   
High-energy ($E\gg 1$ MeV) protons/ions produce a variety of unique nonthermal radiation signatures  \citep{Murphy1997, Share2004, Vilmer2011}, but 
these are not currently detectable at stellar distances \citep{Song2024}.  \citet{Fleishman2022Frontiers} discuss potential signatures in gyrosynchrotron radiation, which is often prominent in stellar flares, that might help to clarify the roles of high-energy ion beams.

Among the current generation of RHD models of stellar flares with electron beam heating \citep{Allred2006, Kowalski2017Broadening, Namekata2020, Kowalski2024}, only the largest heating rates in the low chromosphere and temperature minimum region reproduce the continuum radiation in the NUVA spectra of both flare events on CR Dra.  Much smaller electron-beam flux densities ($\approx 10^{10}-10^{11}$ erg cm$^{-2}$ s$^{-1}$) are widely used to model solar flares \citep[e.g.,][]{Allred2005,Fletcher2008,Kuridze2015,Rubio2016,Carlsson2023,Kerr2024A,Kerr2024B,Simoes2024B}, and the low-energy cutoffs are typically constrained to be $E_c < 40$ keV \citep[e.g.,][]{Veronig2005,WarmuthMann2016,Kleint2016}.  
In Figure \ref{fig:RHD}, we demonstrate that an electron beam with an energy flux density of $5 \times 10^{11}$ erg cm$^{-2}$ s$^{-1}$ and a low-energy cutoff of $25$ keV produces a Balmer continuum spectrum that has the opposite slope as the observations. As expected, a simple optically thin, static slab model calculation in local thermodynamic equilibrium at $T=10,000$ K \citep[dashed red line;][]{Kunkel1970, Kowalski2024LRSP} is even worse.  This clearly demonstrates that the continuum radiation in stellar flares is not as simple as static, uniform, optically thin slab at $T=10,000$ K as assumed by some \citep{Simoes2024} in the solar community.  Other possible sources of very bright NUVA fluxes, such as a hot $T\gtrsim10^7$ K optically thin thermal source, are not plausible hypotheses \citep{HF92}.  \citet{Froning2019} use a semi-empirical hot, dense chromospheric condensation 
to model large blackbody color temperatures in the FUV. However, the densities and temperatures in this model component are not self-consistently produced in  current RHD models. 

The larger inferred  low-energy cutoffs and flux densities of electron beams in M dwarf flares may effectively originate during magnetic reconnection \citep[see discussion in][]{Kowalski2024} and particle acceleration \citep[e.g.,][]{Hamilton1992}.  However, there has been little work on the origin of accelerated particles in stellar atmospheric conditions not typically found in the solar corona.  Alternatively, \citet{Kowalski2023} suggests that similar beam distributions to those with large $E_c$ values could  result from time-dependent transport effects \citep[][see also \citet{Karlicky2012}, \citet{Pechhacker2014}, \citet{Ziebell2022}, \citealt{Annekov2023}]{Kontar2012} in strong magnetic fields, which may saturate the low-coronae of dMe stars \citep{White1994} and inhibit ion-acoustic beam instabilities \citep{LeeBuchner2011, Li2012}.  The beam density of the \texttt{cF13-500-3} model during coronal propagation is $\approx 4 \times 10^8$ cm$^{-3}$, which is well below ambient densities determined from X-ray spectra of dMe stars outside of flaring times \citep{Osten2006, Liefke2008}.  Thus, the Buneman instability and the availability of electrons are not concerns.
Gyrosynchrotron radiation at optically thin radio frequencies \citep[][Tristan et al. in prep]{Osten2016}, ALMA flare emission \citep[which has been observed to be temporally-correlated with a highly impulsive response in the FUV;][]{MacGregor2021}, and linear polarization in $U$-band bursts \citep{Beskin2017} could help to place constraints on the number of relativistic electrons and the strengths of magnetic fields in stellar flares \citep{MacGregor2020}.

\begin{figure}[h]
\centering
\includegraphics[width=1.0\textwidth]{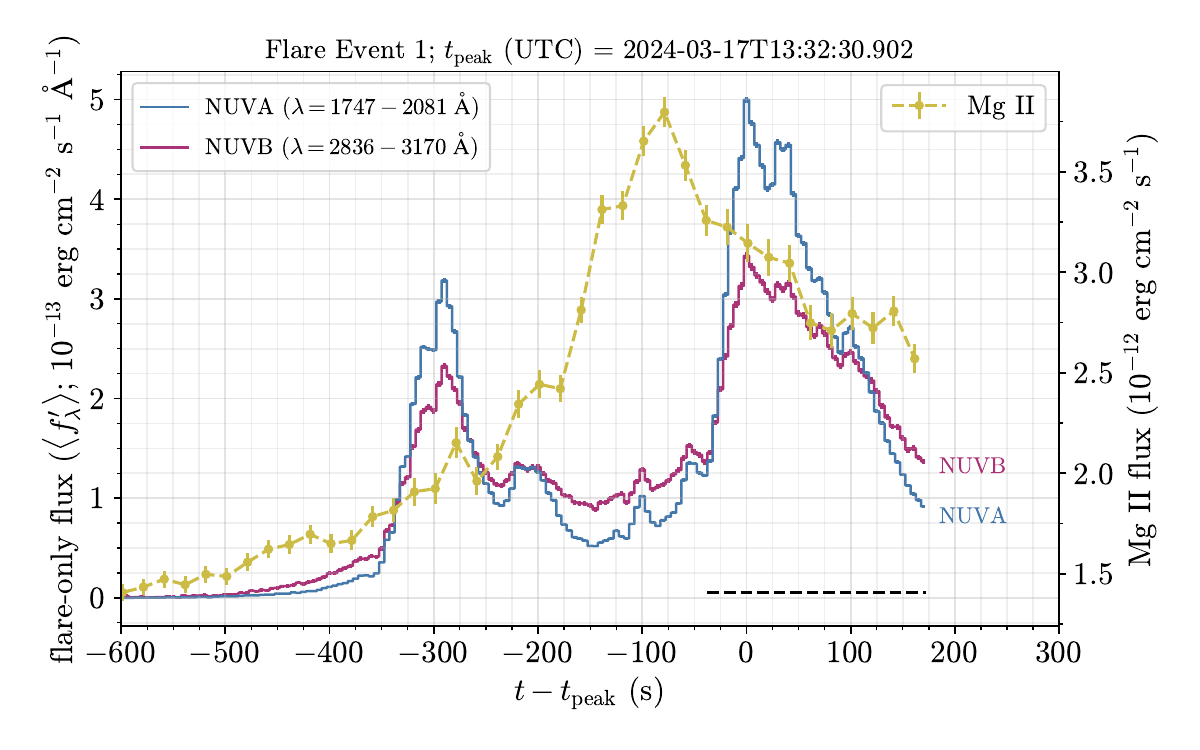}
\includegraphics[width=1.0\textwidth]{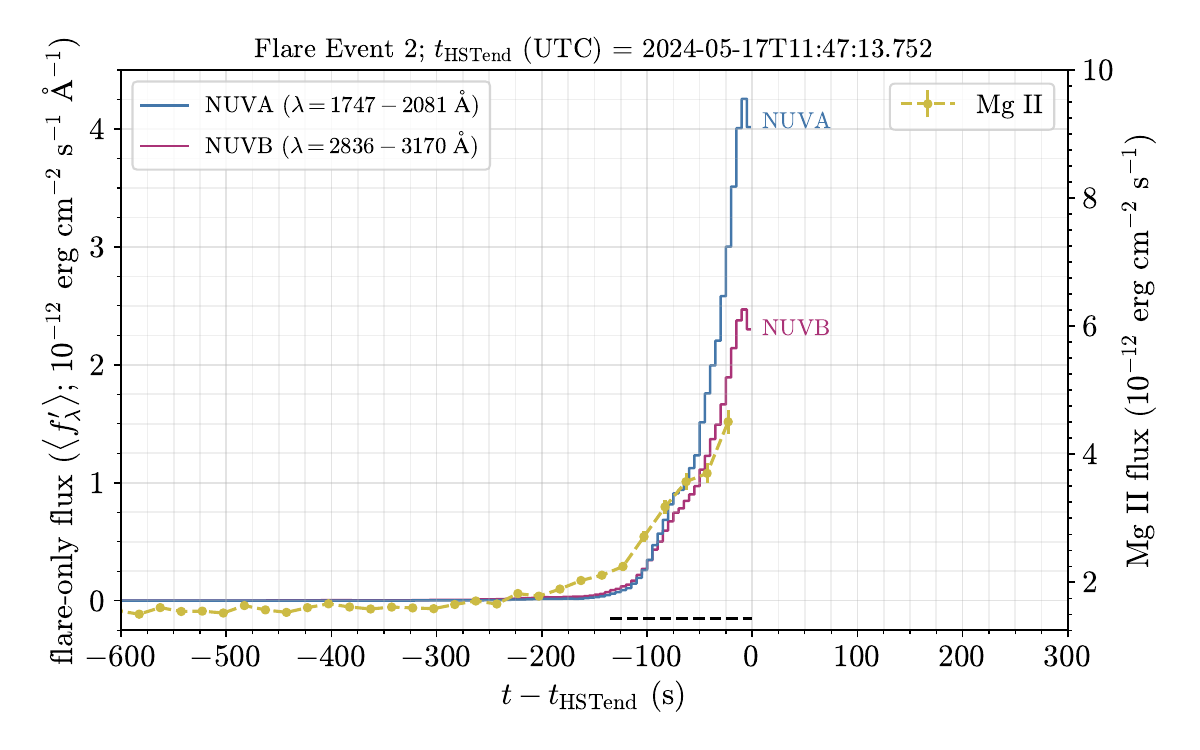}
\caption{Broadband NUV light curves of Flare Event 1 (top) and Flare Event 2 (bottom) at $\Delta t = 5$~s binning. Note the differences in scales on the left y-axes.  The Mg II line fluxes are not pre-flare subtracted and are shown at the 20~s cadence of TESS photometry (Appendix \ref{sec:supplementaryLCs}).  The horizontal dashed lines at the bottom show extraction intervals for the spectra in Figure \ref{fig:spectra}. }\label{fig:lcs}
\end{figure}

We interpret the two-component electron beam model fits  (Section \ref{sec:models}) to represent heterogeneous flaring sources on stars in analogy to chromospheric sources in solar flares.
Notably, the ratio of areas, $\hat{X}_{\rm{low}}/ \hat{X}_{\rm{high}}  = \hat{A}_{\rm{low}}/ \hat{A}_{\rm{high}} \approx 3$, that we infer from the average rise phase spectra of Flare Event 2 is a very similar value to the ratio of the area of bright ribbons to the area of bright kernels in some solar flares \citep[e.g.,][]{Kowalski2022Frontiers}.  Thus, the areas heated by high-energy electrons with high-flux density (\texttt{cF13-500-3}) could represent concentrated heating in kernels, which are surrounded by extended ribbons that are heated by lower-energy electron beams (\texttt{m2F12-37-2.5})\footnote{This dichotomy, however, is not obviously a stellar analogue of a ``core-halo'' morphology that is discussed in optical observations of solar flares \citep{Neidig1993, Isobe2007, Namekata2022}.  It is thought that ``halo'' structures around bright kernels could be due to radiative-backwarming of the photosphere \citep{Allred2006, Fisher2012}, which causes H$^-$ emission to increase.  H$^{-}$ opacity contributes very little to the wavelengths in the HST/NUV range \citep{Garcia2005} and in the slit-jaw 2832 images \citep{Kleint2016} from the Interface Region Imaging Spectrograph \citep[IRIS;][]{DePontieu2014} that show the extended ribbons around the concentrated kernels \citep{Kowalski2022Frontiers}. }.   The impulsiveness calculations from the light curves in NUVA and NUVB (Section \ref{sec:analysis}; Figure \ref{fig:lcs}) could further support a model framework wherein two rather distinct heating components with different temporal evolution contribute to the total flare continuum flux; within the solar analogy, the kernels are transient while the ribbons are persistent.  The inferred flare areas, $\hat{A}_{\rm{high}} \approx 2 \times 10^{19}$ cm$^{-2}$ and  $\hat{A}_{\rm{low}}\approx 6.5\times 10^{19}$ cm$^{2}$, are larger in the CR Dra stellar megaflares than the total areas directly measured in solar flare kernels and ribbons, which typically range from $10^{16} - 10^{18}$ cm$^2$ but sometimes extend up to $\approx 2 \times 10^{19}$ cm$^2$ at certain wavelengths \citep{Neidig1994, Fletcher2007}.  As additional comparison, the inferred flare areas in the Great Flare of AD Leo with the \texttt{mF13-500-3} and the \texttt{m2F12-37-2.5} models \citep{Kowalski2022Frontiers} are factors of ten and two smaller than the respective RHD components (\texttt{cF13-500-3} and \texttt{m2F12-37-2.5}) that are fit to Flare Event 2 on CR Dra. The CR Dra megaflare areas are yet only a small fraction of the visible stellar hemisphere, $\hat{X} \approx 0.01$, which implies compact sources.  

The spatial resolution of solar data constrains how spectral and temporal heterogeneity across chromospheric flare sources  contribute to Sun-as-a-star signals \citep{Namekata2022, Pietrow2024}. \citet{Namekata2022} demonstrate that the brightest and broadest H$\alpha$ spectra differ markedly from spatially integrated spectra, which more closely resemble spectra from the diffuse regions with weaker emission line intensity.  Ostensibly, this supports superposing RHD model spectra to represent heterogeneous stellar flare sources.  Similar comparisons of solar data from IRIS to other flares in the HST Treasury program that are similar in energy to large solar flares would be valuable for constraining the amount of heterogeneities in UV sources.  For example, it would be valuable to determine if any very bright kernels in solar flares \citep[e.g.,][]{Jess2008, Krucker2011} exhibit a brighter and more impulsive FUV continuum source than expected, and, further, if these sources also have a gradual response in Mg II $h$ and $k$. \citet{Tian2015} discuss a nearly simultaneous evolution of the intensities of Mg II, the NUV continuum, and the FUV continuum in the impulsive phase of a solar flare.  In another flare, the FUV continuum intensity in the bright ribbons appears to exhibit a more prominent spike phase than the NUV continuum intensity, which peaks after the wavelength-integrated Mg II $k$ light curve.  To our knowledge, these differences have not been previously discussed and interpreted in terms of solar flare heating mechanisms, perhaps because it is generally thought that irradiation of the low atmosphere by transition region lines adequately explains the FUV continuum response \citep{Machado1982, Doyle1992B, Young2015, Tian2015}.  Solar Dynamics Observatory/Atmospheric Imaging Assembly (SDO/AIA) 1700 \AA\ \citep{Lemen2012} data may facilitate closer comparisons than the IRIS/FUV spectra to the rising NUVA spectra in stellar flares, but these data lack spectral information and saturate in the impulsive phase of large solar flares.

The relatively gradual response of the Mg II emission lines in the new NUV spectra (Figure \ref{fig:lcs}(top)) may indicate that the most prominent sources of chromospheric heating and continuum formation in stellar flares differ from bright kernels \citep[e.g.,][]{Tian2015,Kleint2016} in solar flares.  \citet{Kowalski2019HST} compare the Mg II line and NUV continuum ($\lambda \approx 2650$ \AA) fluxes in two lower energy flares.  Two events in their sample show longer $t_{1/2}$ values, and one shows a delay of the peak Mg II line flux at 60~s cadence.  \cite{HP91} have previously discussed similarities to the Ca II H and K line fluxes, which show delayed light curve peaks, in the gradual decay phase of the 1985 Apr 12 Great Flare of AD Leo.  If Mg II $h$ and $k$ respond very gradually and peak after the continuum and hydrogen Balmer lines like the resonance lines of Ca II H and K are widely observed to do in other dMe events \citep{Garcia2002, Kowalski2013}, then the light curve evolution in the top panel of Figure \ref{fig:lcs} is expected. To date, no self-consistent  explanation exists for the delayed peaks of the Mg II and Ca II resonance lines in stellar flares.   However, contributions from white-light ``post flare'' loops \citep{Heinzel2018, Yang2023} and radiative backheating of large regions of the chromosphere \citep{HF92,Hawley2003,Fisher2012} surrounding the extended beam-heated ribbons are tantalizing possibilities.

To further argue that extreme heating to $T \gtrsim 10,000$ K occurs in the deep stellar atmosphere where large continuum optical depths can develop \citep[e.g.,][]{Kowalski2023}, we show a suggestive resemblance of the HST/COS SED of Flare Event 2 to the observed UV spectrum of Vega \citep{Bohlin2014, Bohlin2014B} in Figure \ref{fig:summaryfig}.  \citet{Kowalski2011IAUS} find that the newly-formed flare optical spectra of secondary events in the decay phase of an M-dwarf megaflare resemble the spectrum of Vega.  \citet{Kowalski2017Broadening} modeled these flare spectra also with heating from an electron beam with $E_c=500$ keV, but the energy flux density is a factor of five smaller than the \texttt{cF13-500-3} model.  In Figure \ref{fig:summaryfig}, the UV spectrum of Vega bears a few interesting similarities to the flare spectra of CR Dra:  the overall shape of the Vega  spectrum in the NUVB range is relatively flat, and there is a  spectral break that is followed by a steeper increase at shorter wavelengths starting at $\lambda \approx 2400$ \AA.  The break in Vega's spectrum is due to a confluence of Fe II bound-bound transitions \citep{Garcia2005};  we speculate that spectral syntheses of the relativistic electron-beam heated model atmospheres with Fe II opacity may produce flatter NUVB shapes that are closer to the observations.  This additional source of NUV opacity may improve the RHD model fits, which currently over-predict the observed TESS flux in Flare Event 2.

The NUV flare spectra have important applications beyond stellar flare physics.
The spectra improve the empirical and phenomenological (blackbody) interpretations of broadband UV flare data \citep{Robinson2005, Brasseur2023, Tristan2023, Paudel2024, Berger2024}, which are more readily obtained than spectra.  There are several upcoming missions that will have one or two broadband UV imaging capabilities with stated goals of characterizing stellar activity.  The Star-Planet Activity Research CubeSat \citep[SPARCS;][]{SPARCS1} will have two bandpasses at $\lambda = 1530-1710$ \AA\ and $\lambda =2580-3080$ \AA, the Quick Ultra-Violet Kilonova surveyor \citep[QUVIK;][]{QUVIK1,QUVIK2} will have two bands at $\lambda = 1400-1900$ \AA\ and $\lambda = 2600-3600$ \AA, ULTRASAT \citep{ULTRASAT} will have one band at $\lambda = 2200-2900$ \AA, and the Ultraviolet Explorer \citep[UVEX;][]{UVEX}
will have two bandpasses at $\lambda = 1390 - 1900$ \AA\ and $\lambda = 2030-2700$ \AA. With just one or two bandpasses, it is not possible to constrain a spectral superposition with two RHD models or two blackbody temperatures; thus additional constraints will be necessary for accurate modeling of stellar flaring activity.  Other, related applications of the new NUV flare spectra pertain to exoplanet habitability experiments \citep{Abrevaya2020} and photochemical simulations. 
The FUV radiation and short-wavelength NUV photons photolyze atmospheric carbon dioxide, methane, oxygen (O$_2$) and water, among other gases.  Thus, the correct account of the flux in this wavelength region is fundamental to predict the effect of flares in planetary atmospheric chemistry and prebiotic chemistry  \citep{Segura2010, Tian2014,Loyd2018, Schwieterman2019,Chen2021}.  For example, more FUV  would produce more O$_3$ in a O$_2$-rich atmosphere, while in a CO$_2$-rich atmosphere more CO and O$_2$ would be created and CH$_4$ photolysis could lead to haze formation.

\section{Summary \& Conclusions} \label{sec:conclusions}
It has generally been assumed that a $T\approx 10^4$ K blackbody model is a sufficiently accurate representation for M dwarf flare radiation, but there has never been a precise test of the predicted peak flux with NUV spectra.  We present such spectra in the impulsive phase of two megaflare events, which provide the first constraints on what happens spanning $\Delta \lambda \approx 1500$ \AA\ across the NUV.  The long-wavelength NUV spectra have blackbody color temperatures of $T \approx 10^4$ K, in line with general expectations from longer wavelengths within the $U$ band \citep{Fuhrmeister2008}.  However, the unassuming (due to a lack of commonly studied resonance lines) wavelength range  from $1700-2100$ \AA\  varies more impulsively and exhibits larger peak continuum fluxes and hotter blackbody color temperatures.  These unprecedented findings suggest that the peak of the white-light continuum radiation is located at $\lambda \lesssim 1700$ \AA\ and a break to a rising spectrum occurs between $\lambda \approx 2100$ \AA\ and $\approx 2700$ \AA.  For the first time, we have shown time-resolved evolution of the Mg II emission line flux in the impulsive phase of a stellar flare.  The wavelength-integrated Mg II emission line flux evolution is almost completely decoupled from the time-evolution of the continuum radiation, which dominates the integrated energy in the NUV.

These spectral properties constrain a remarkable amount of impulsive-phase heating to the deep stellar atmosphere.
Optically thin hydrogen recombination radiation has been used to model solar flare continuum intensities in the IRIS/NUV \citep[e.g.,][]{Heinzel2014, Kleint2016, Kowalski2017Mar29}, which overlaps with our NUVB range, and continuum intensities around 2000 \AA\ \citep{Dominique2018}, which overlaps with our NUVA range.  Considered together, the new short- and long-wavelength NUV spectra provide the most compelling evidence to date that this paradigm does not explain flaring M-dwarf continuum radiation \citep[even with a modestly heated upper photosphere;][]{Neidig1993, Allred2006, Kleint2016, Kowalski2017Mar29}, an idea that was pioneered by \citet{Kunkel1970} and \citet{HF92}.

Among all current stellar flare RHD models, the one that best reproduces the continuum radiation in the new spectra at $\lambda = 1700-2100$ \AA\ has very large heating rates from relativistic electron beams in the deep atmosphere, which result in hot blackbody-like (i.e., originating from large and wavelength-dependent continuum optical depths) thermal radiation.  A superposition of two model radiative flux spectra, which may plausibly represent bright spatially extended ribbons and compact kernels, is a semi-empirical explanation for two blackbody temperatures across the NUV. However, the particle beam energies in each model component are much larger than canonical values in solar flares.
The consequences of large flux densities of relativistic electron beams in stellar flares should be taken seriously in the absence of evidence of high-energy proton beams or another physical explanation for the NUV spectra.

\begin{figure}[h]
\centering
\includegraphics[width=1.0\textwidth]{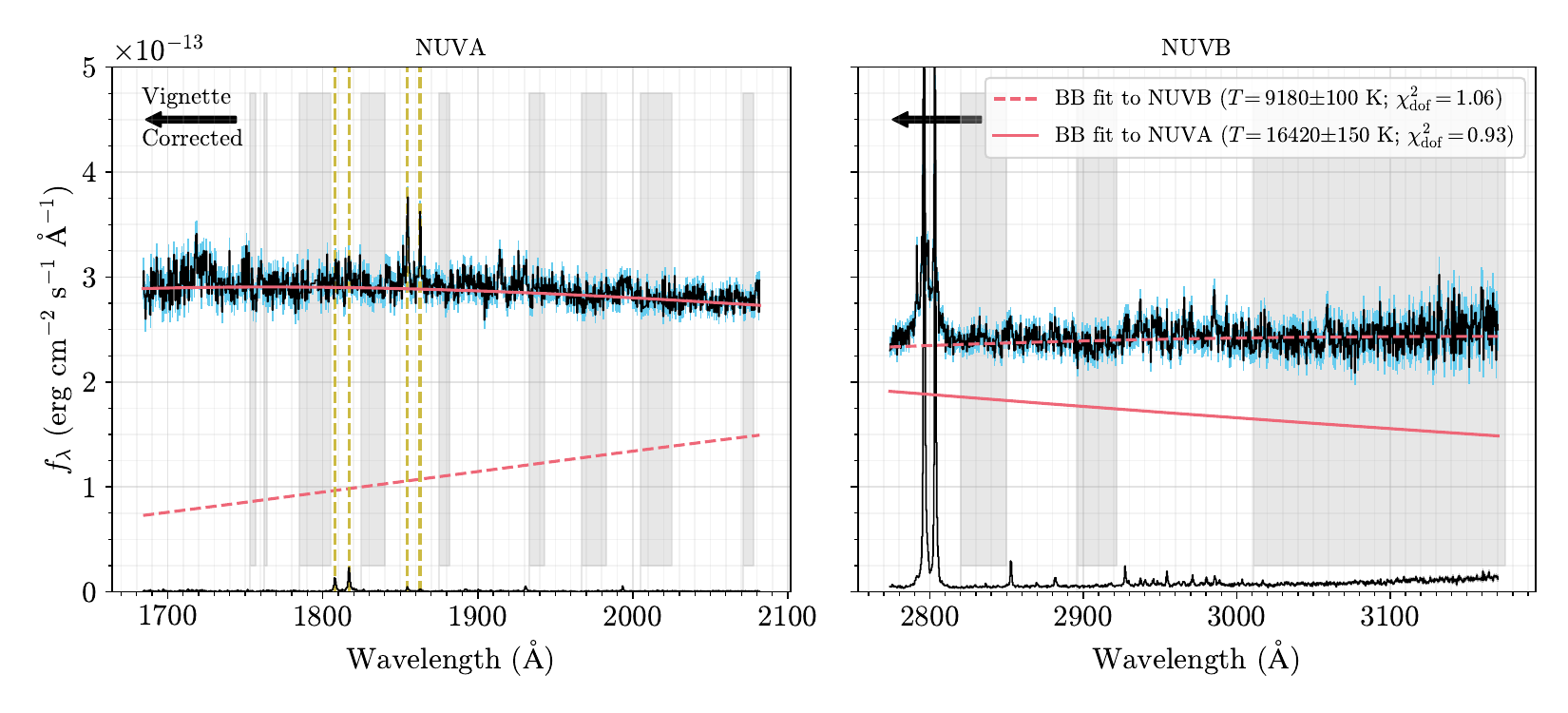}
\includegraphics[width=1.0\textwidth]{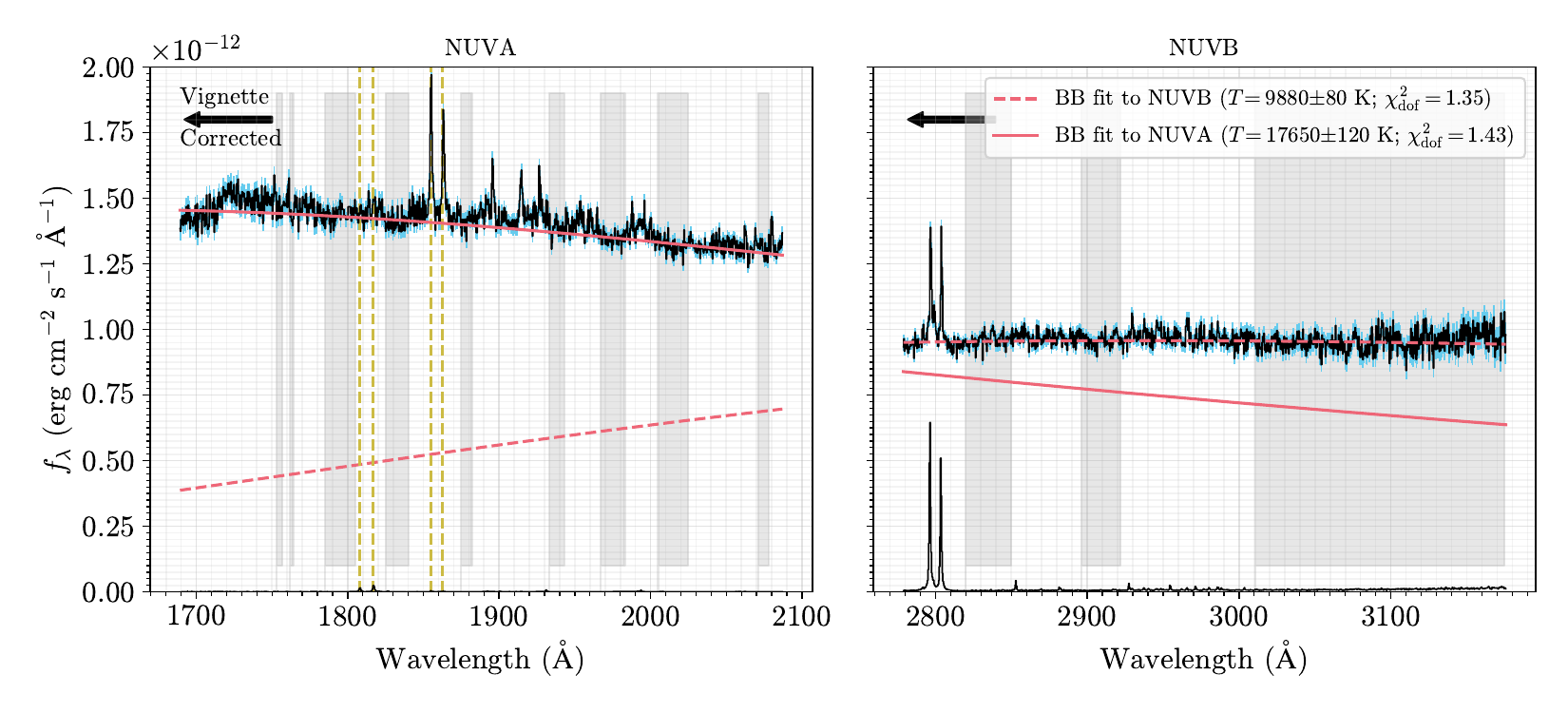}
\caption{ Impulsive-phase, flare-only NUV spectra of Flare Event 1 (top; UTC 2024-03-17 13:31:53.40 - 13:35:23.40) and Flare Event 2 (bottom; UTC 2024-05-17 11:44:58.75 $-$ 11:47:13.75).   The times over which the photons are summed within each flare event are indicated with horizontal dashed lines in Figure \ref{fig:lcs}.  Blackbody functions are fit to the gray shaded wavelength regions within each stripe, and they are extrapolated to the other stripe.  The preflare spectra are obtained by averaging over all of the quiescent time before each flare event (UTC 2024-03-17 13:05:18.40 $-$ 13:22:58.40 before Flare Event 1 and UTC 2024-05-17 11:18:18.752 $-$ 11:36:38.752 before Flare Event 2).  Note the difference between the flux scales on the left axes. The yellow dashed lines indicate rest wavelengths of Si II and Al III (see text).  Left-facing arrows indicate the regions within each stripe that are affected by and corrected for vignetting (Appendix \ref{sec:hstdata}).
 }\label{fig:spectra}
\end{figure}

    \begin{figure}
\includegraphics[width=1.0\textwidth]{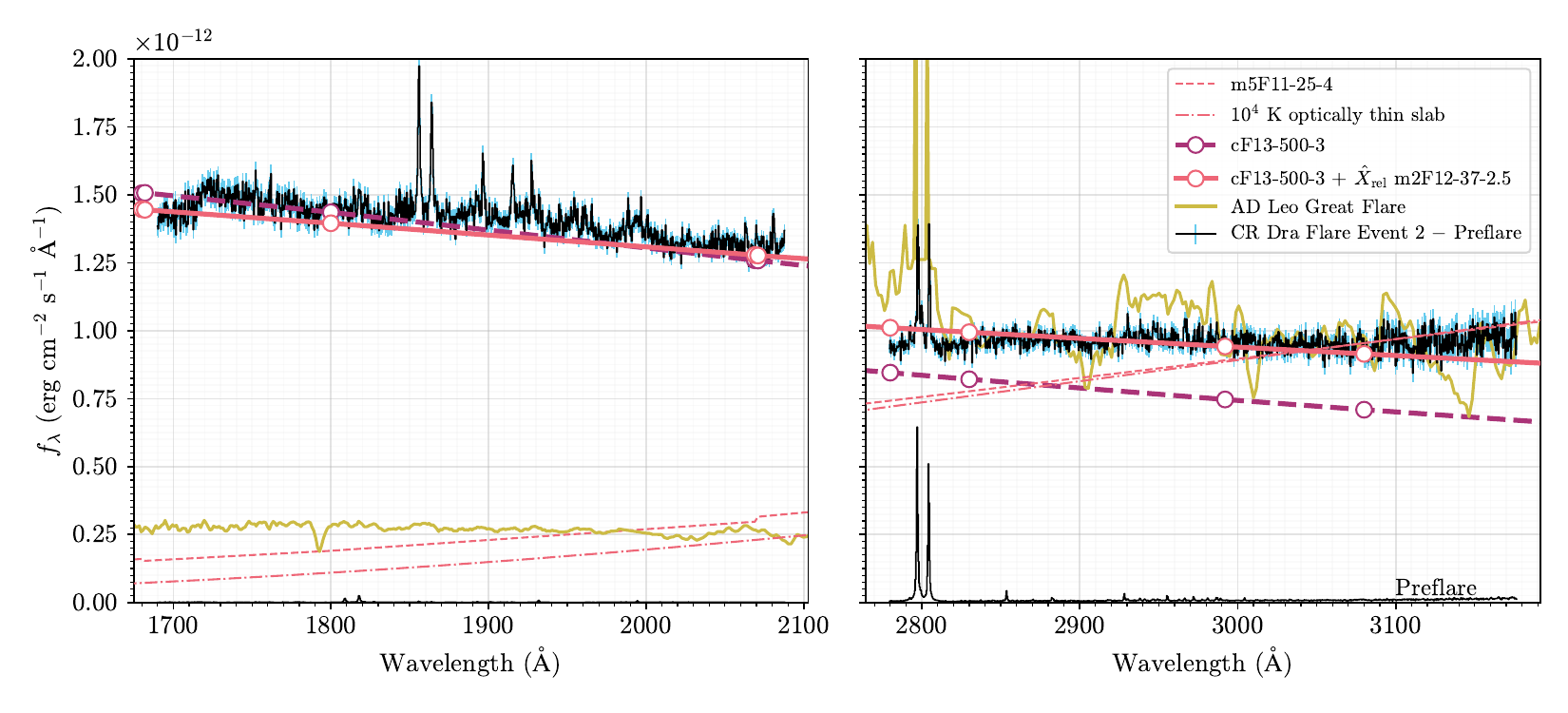}
\caption{ RHD model superposition that is fit to the rise-phase spectra of Flare Event 2 from the bottom panel of Figure \ref{fig:spectra}.   A model with $E_c=500$ keV, $\delta = 3$ and an energy flux density of $10^{13}$ erg cm$^{-2}$ s$^{-1}$ (\texttt{cF13-500-3}) is also shown without any second RHD model component;  this is scaled to the continuum wavelengths around $\lambda = 1800$ \AA\ in  NUVA.  The AD Leo Great Flare spectral template  is shown in yellow and is scaled to the NUVB spectral region;  note that these IUE spectra have a resolving power of $R \approx 300$, compared to $R \approx 3000$ for the HST/COS data.  A lower-flux RHD model (\texttt{m5F11-25-4}) with electron beam parameters inferred in a major solar flare \citep{Kowalski2017Mar29}  and an optically thin slab model calculation produce decreasing continuum distributions that are opposite of the observed flare SEDs.
 }\label{fig:RHD}
    \end{figure}

\begin{figure}[h!]
\centering
\includegraphics[width=0.5\textwidth]{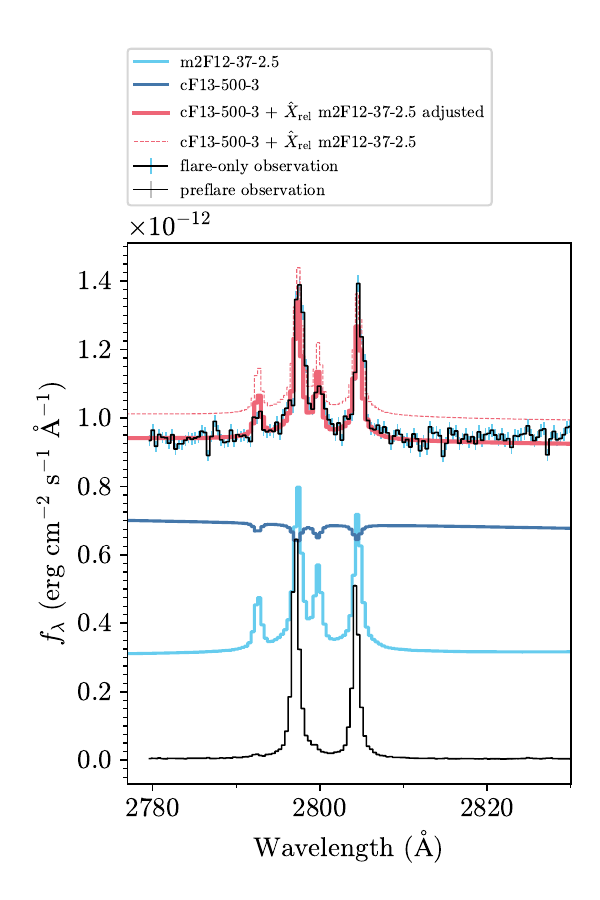}
\caption{ Mg II calculations from the RH code compared to the rise-phase, flare-only spectrum (thick black line; UTC 2024-05-17 11:44:58.75 $-$ 11:47:13.75) of Flare Event 2.  Thin black line:  pre-flare observation;  dark blue: model component 1 (\texttt{cF13-500-3} at $t=2.2$~s);  light blue:  model component 2 (\texttt{m2F12-37-2.5} at $t=1$~s);  dashed red:  total model fit to continuum regions;  thick solid red: total model spectrum scaled by 0.93. Note, this spectral region may suffer from residuals from our vignetting correction (Appendix \ref{sec:hstdata}).
 }\label{fig:MgII}
\end{figure}

\begin{figure}[h!]
\centering
\includegraphics[width=1.0\textwidth]{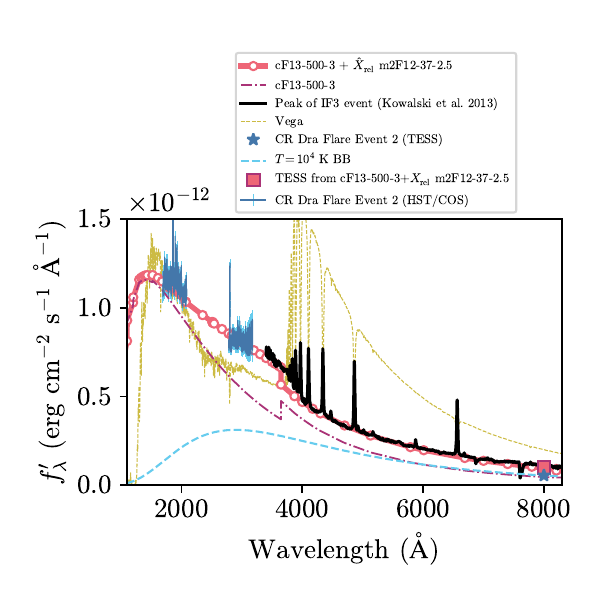}
\caption{ Expanded view ($\lambda = 1100-8300$ \AA) of the HST/COS spectra and TESS data from the early rise phase of Flare Event 2 at UTC 2024-05-17 11:45:01 $-$ 11:47:01.  In the RHD model combination, we calculate that $\hat{X}_{\rm{rel}} = 4.5$ (see text). The red square symbol is the flux ($5.5 \times 10^{-14}$ erg cm$^{-2}$ s$^{-1}$ \AA$^{-1}$) of the total RHD model weighted by the TESS bandpass, plotted at $\lambda = 8000$ \AA.  The peak optical flare spectrum from the giant IF3 event on YZ CMi is scaled by the flux at $\lambda = 4170$ \AA\  so that it overlaps with the RHD model flux. The YZ CMi flare spectrum should be viewed as a rough \emph{prediction} for the optical and Balmer jump properties that are unobserved in the CR Dra impulsive phase.   The CALSPEC observation of Vega is shown with arbitrary scaling to the NUVA;  Vega's spectrum demonstrates that there is a well-understood astrophysical environment that produces a steep rise from the NUVB range to the NUVA.  }\label{fig:summaryfig}
\end{figure}

\clearpage

\begin{acknowledgements}

We thank an anonymous referee for helpful comments.  AFK acknowledges funding support from HST GO 17464 and helpful comments on the paper from Dr. John P. Wisniewski.
YN acknowledges funding from NASA ADAP 80NSSC21K0632, NASA TESS Cycle 6  80NSSC24K0493, NASA NICER Cycle 6 80NSSC24K1194, and HST GO 17464.  We thank the Program Coordinator Bill Januszewski for help planning the HST observations.  AFK thanks Dr. Joel C. Allred for the updating the \texttt{RADYN} code to the \texttt{RADYN+FP} code thus allowing the proton beam simulation, Dr. Mats Carlsson for the use of the \texttt{RADYN} code, Dr. Suzanne L. Hawley for helpful conversations about near-ultraviolet flare spectra, Dr. Enoto Teruaki for helpful discussions about an early draft, Clara Brasseur for help on proposal preparation and discussions, Dr. Ward Howard for discussions on the timestamps in TESS headers, Elena Mamonova for helpful discussions, and Dr. Adina Feinstein for helpful discussions about FUV spectra.  We make use of Paul Tol's color tables (\url{https://personal.sron.nl/~pault/}).

Some of the data presented in this article were obtained from the Mikulski Archive for Space Telescopes (MAST) at the Space Telescope Science Institute.  The specific HST/COS and TESS observations analyzed can be accessed via \dataset[https://doi.org/10.17909/dbr7-3f98]{https://doi.org/10.17909/dbr7-3f98}. STScI is operated by the Association of Universities for Research in Astronomy, Inc., under NASA contract NAS5–26555. Support to MAST for these data is provided by the NASA Office of Space Science via grant NAG5–7584 and by other grants and contracts. Funding for the TESS mission is provided by the NASA Explorer Program. 
The NICER data used in this study were obtained through the GO program Cycle 6 ID: 7055.
NICER analysis software and data calibration were provided by the NASA NICER mission and the Astrophysics Explorers Program.
This research also makes use of observations from the Las Cumbres Observatory (LCO) global telescope network. We used the LCO observation time allocated to the University of Colorado.

\end{acknowledgements}

\appendix    

\section{Data Reduction} \label{sec:data_appendix}

\subsection{HST/COS Data} \label{sec:hstdata}

We obtain the COS data from the Mikulski Archive for Space Telescopes (MAST).  We use \texttt{costools} to split the \texttt{corrtag} files into intervals of 5~s, intervals of 20~s that correspond to the cadence of the TESS data (Section \ref{sec:tessdata}), and intervals that correspond to interesting times over the flares.  We also choose a long interval within the same orbit but before each flare to sum into a quiescent spectrum, and these are subtracted to give flare-only spectra ($f^{\prime}_{\lambda}$).  The spectra are extracted with \texttt{calcos} using a 21 pixel boxcar aperture about the default y-pixel location of each stripe (which are well within a fraction of a pixel from those found by centroiding).  The 21 pixel aperture is determined by inspecting the spatial profiles of the \texttt{flt} images produced by \texttt{calcos}.  The aperture size choice considers a balance between signal-to-noise and signal (ISR COS 2017-03).  We add $0.33$ \AA\ and $-0.76$ \AA\ to the wavelength calibration of the two orbits, respectively;   Gaussian-centroiding the Mg II $h$ and $k$ emission lines in the preflare spectra determines these adjustments, which are within the pipeline accuracy of 175 km s$^{-1}$ for the COS/NUV (Section 5.1.11 of the COS Instrument Handbook; ISR COS 2024-07) and the variable radial velocities of the components of CR Dra \citep[$-10$ to $-50$ km s$^{-1}$;][]{Sperauskas2019}. \texttt{calcos} applies a flux calibration and the most recent time-dependent sensitivity correction \texttt{8782023sl\_tds} (COS STAN July 2024).

Two further steps are taken to accurately flux calibrate the spectra.  First, an aperture correction is applied to the spectra as follows.  We divide spectral extractions using the default wide aperture of 57 pixels by a corresponding 21 pixel extraction.  A constant correction of 1.083 is determined for NUVB, but a wavelength-dependent linear correction is needed for NUVA.  The corrections depend relatively weakly on wavelength and range linearly from 1.115 at the longest wavelength to 1.151 at the shortest wavelength in NUVA\footnote{We fit linear functions to the ratios of the 57 to the 21 pixel extractions.  The best fit slopes and intercepts are $m=-7.208\pm1.156\times10^{-5}$ and $b=1.268\pm0.022$ for NUVA and $m=-3.15\pm11.34 \times 10^{-6}$ and $b=1.092 \pm 0.033$ for NUVB.}.
We check the consistency of these corrections with orbit-integrated spectra and with spectra around the peak of Flare Event 1.
Second, the shortest wavelengths of each stripe in the NUV suffer from reduced signal due to a detector shadow (vignetting; Section 5.1.12 of the COS Instrument Handbook). Because a correction is not applied by the \texttt{calcos} pipeline,  we determine one by comparing the fluxes at overlapping wavelengths in COS/G230L observations (GO 17319) of WD1057+719 with 2635 \AA\ and 2950 \AA\ central wavelengths.   We use these data to apply a linear correction ranging from 1.230 to 1.0 in NUVB and 1.196 to 1.0 in NUVA over $\approx 110$ pixels that are flagged with data quality flag $=4$ (indicated by arrows in Figure \ref{fig:spectra}).  For another target (GJ 1243) in GO 17464, we find that the continuum in the preflare spectra aligns with overlapping regions of a quiescent spectrum in data from 2014 that were obtained with G230L and a central wavelength of 2635 \AA\ \citep{Kowalski2019HST}.  After the correction, there still remains a little bit of a decreasing slope indicative of the vignetting in some of the brighter flare spectra.  We suggest that the Space Telescope Science Institute provide a formal correction to the shadowed region of NUV data, which we show could be salvaged from the MAST.
The shortest $\approx$ 40 \AA\ in each stripe are most heavily affected by vignetting and are not included in the blackbody temperature fits (Section \ref{sec:analysis}; Figure \ref{fig:spectra}).   The vignetting and aperture corrections do not affect the conclusions of this paper.

We use \texttt{astropy}'s \texttt{time} module to convert the times of the HST data (which are given in MJD format on the UT scale, assumed to be the UTC scale; \citealt{Eastman2010}) to the BJD format on the Barycentric Dynamical Time (TDB) scale.  We then add the light travel time to the barycenter using the stellar coordinates and \texttt{time}'s submodule \texttt{light\_travel\_time} (following \texttt{astropy}'s online documentation).  This transforms the mid-exposure times of the HST data to the time system in the TESS data (Appendix \ref{sec:tessdata}) headers.

\subsection{TESS Data} \label{sec:tessdata}
  We retrieve short-cadence ($\Delta t = 20$~s) \texttt{SAP\_FLUX} data from the MAST.  The TESS data provide contextual information about the broadband $\lambda = 6000-10,000$ \AA\ response during the NUV flare events.  Since the HST observations ended during each event, the TESS data are used to calculate the flare energies and durations (Appendix \ref{sec:supplementaryLCs}).

\subsection{NICER Data} \label{sec:nicer_data}

The Neutron Star Interior Composition ExploreR (NICER; \citealt{Gendreau2016}) observed on 2024 March 17 and 2024 May 17 for NICER Guest Observer (GO) Program 7055. NICER made 8 observations of the soft X-ray (0.2--12 keV) on each date, and each exposure is $\approx$2 ks.

We follow the standard NICER data analysis procedures \citep[e.g.][]{Inoue2024_PASJ}.
We retrieved the data from OBSIDs 7555020101 and 7555020203 from the HEASARC archive, and we used \texttt{nicerl2} in HEASoft version 6.33.2 to 
filter and calibrate the raw data using the calibration database (\texttt{CALDB}) version \texttt{xti20240206}. 
Filtered data were barycenter-corrected 
using \texttt{barycorr} at the target position of (RA, Dec) = (244.2723, 55.269103).   Then we extracted light curves from the filtered and barycenter-corrected event file with \texttt{xselect}. We also generated source and background spectra with \texttt{nibackgen3C50} (\citealt{Remillard2022_AJ}).   Response files (\texttt{RMF} and \texttt{ARF}) were generated  using \texttt{nicerrmf} and \texttt{nicerarf}.  Light curves and 1$\sigma$ error bars are extracted at $\Delta t = 64$~s within each exposure.  Here we only confirmed that the extracted spectra have no features affecting reliability of the $0.3-4.0$ keV light curve discussion (Section \ref{sec:supplementaryLCs}).  The background count rates of $\approx0.65$ and 0.83 counts s$^{-1}$ for Flare Event 1 and 2, respectively, are not subtracted from the light curves. 

\subsection{LCO Data} \label{sec:LCO_data}

Optical $V$-band photometry observations of CR Dra were conducted on 2024 March 17 using Las Cumbres Observatory (LCO) 0.4m telescopes with the QHY600 CMOS cameras (\citealt{Brown2013_PASP,Harbeck2024}). The exposure time is 4~s. The data are reduced with the LCOGT automatic pipeline BANZAI\footnote{\url{https://github.com/LCOGT/banzai}}, which masks bad-pixels, applies an astrometric solution, and performs bias and dark subtraction. We use AstroimageJ (\citealt{Collins2017_AJ}) to perform aperture photometry with several nearby reference stars.  The $V$-band data (with times in UTC) are plotted relative to the UTC of the peak of the flares in the HST data.

\section{Supplementary Light Curves and Calculated Quantities} \label{sec:supplementaryLCs}

\begin{figure}[h!]
\centering
\includegraphics[width=1.25\textwidth,angle=90]{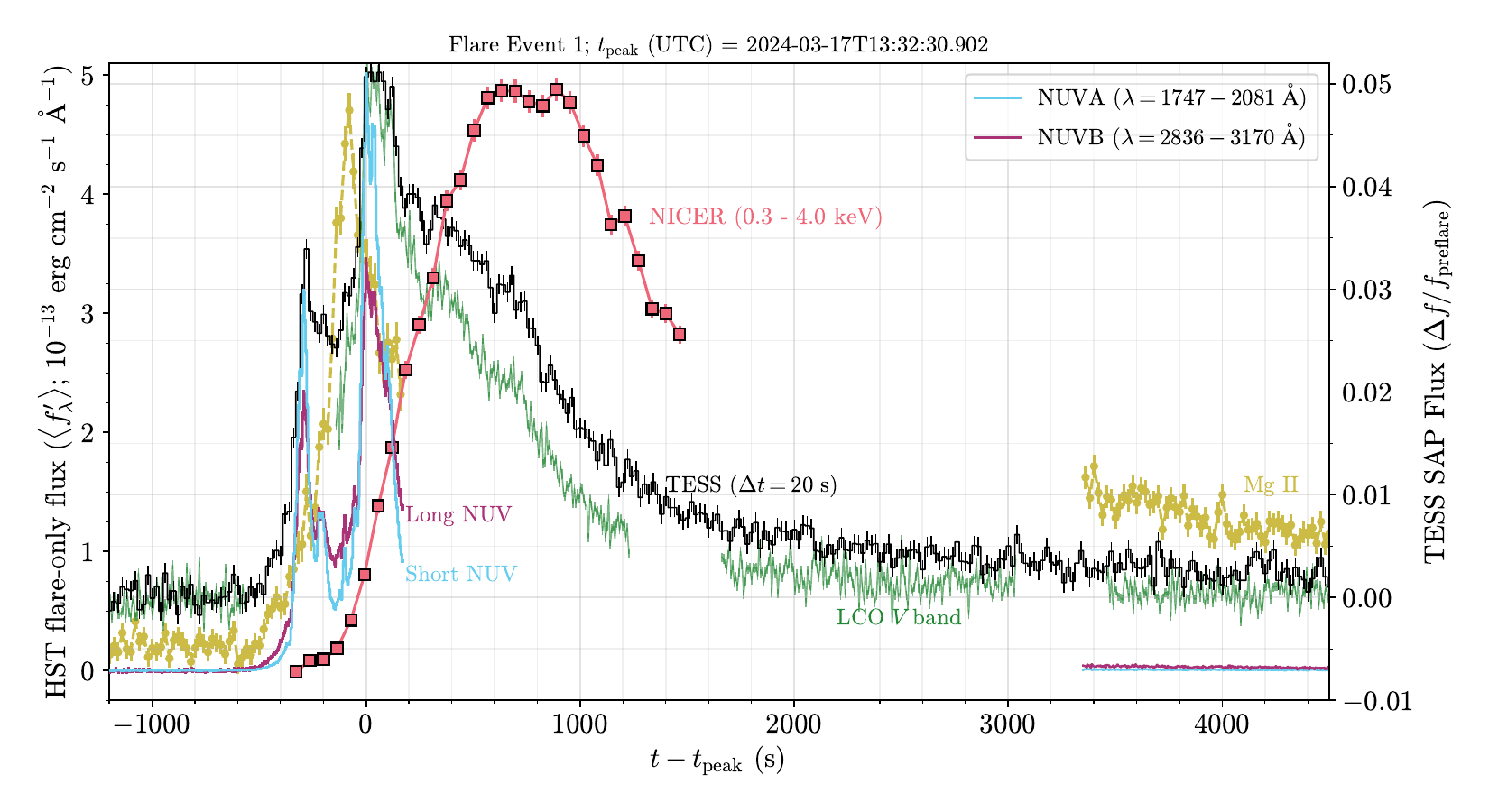}
\caption{ Light curves of Flare Event 1.  The axes for the $E=0.3-4$ keV X-ray (light red, squares), Mg II (yellow), and $V$-band (green) data are not shown for clarity.  The X-ray flux axis runs from $4.5$ to $55$ count s$^{-1}$, the continuum-subtracted Mg II line flux axis runs from $1.25-4.00 \times 10^{-12}$ erg cm$^{-2}$ s$^{-1}$, and the $V$-band axis is relative flux and corresponds to $\Delta f / f_{\rm{preflare}} = -0.05 - 0.275$.  The cadence of the X-ray data is $\Delta t = 64$~s.
 }\label{fig:lc_flare1}
\end{figure}

\begin{figure}[h!]
\centering
\includegraphics[width=1.25\textwidth,angle=90]{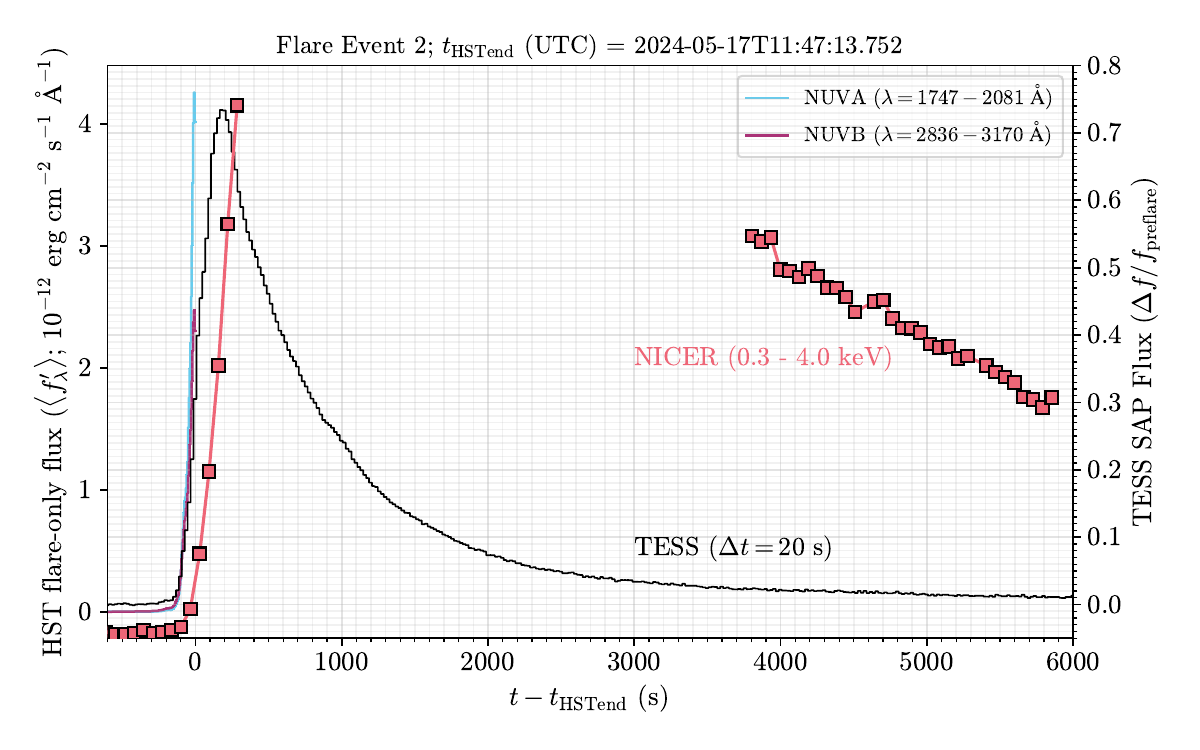}
\caption{ Light curves of Flare Event 2.  The axis for the X-ray (light red, squares) light curve is not shown for clarity.  The X-ray flux axis runs from $4.5$ to $100$ count s$^{-1}$ and the cadence is $\Delta t = 64$~s.
 }\label{fig:lc_flare2}
\end{figure}

In this appendix, we show supplementary light curve data of Flare Event 1 (Figure \ref{fig:lc_flare1}) and Flare Event 2 (Figure \ref{fig:lc_flare2}), and we describe the calculation of energies from the TESS \texttt{SAP\_FLUX} data.   The additional light curve data provide important contextual information for the HST/COS light curves that are discussed in the main text.

The supplementary light curves indicate that the two NUV flare events are associated with luminous soft X-ray flares.  
The $E=0.3-4.0$ keV X-ray light curve data (Figure \ref{fig:lc_flare1}; Appendix \ref{sec:nicer_data}) during Flare Event 1 shows a typical slow rise and late peak after the TESS peak. The $E=0.3-4.0$ keV X-ray light curve of Flare Event 2 (Figure \ref{fig:lc_flare2}) extends just past the peak of the TESS light curve.  After about an hour into the flare, the X-ray flux is decaying but is still elevated by just as much as near the peak of the TESS light curve.   The X-ray peak delays with respect to the NUV light curve peaks are qualitatively consistent with the Neupert effect \citep{Neupert1968}, which is widely reported in solar \citep[e.g.,][]{Dennis1993, Veronig2002} and stellar \citep{Hawley1995, Gudel1996, Gudel2002, Osten2004, Fuhrmeister2011,Lalitha2013,Cabellero2015, Osten2016, Tristan2023} flares.  Detailed analyses of the X-ray data will be presented in a future paper (Notsu et al 2025, in prep). 

The TESS data show the red-optical and NIR extension of the white-light response in each flare, and they are presumably dominated by continuum emission \citep{HP91, Schmidt2007, Fuhrmeister2008, Kowalski2013}.
The TESS light curve of Flare Event 1 (Figure \ref{fig:lc_flare1}) shows that the HST orbit ends at the termination of the fast decay phase.  At the start of the next orbit, the  Mg II line emission is still elevated and continues to decay.  The $V$-band light curve data (Appendix \ref{sec:LCO_data}) reaches a peak $\Delta f / f_{\rm{preflare}} \approx 0.27$ and has a slightly faster decay than the TESS light curve, which peaks at $\Delta f / f_{\rm{preflare}} \approx 0.05$.  We note similar low-amplitude variations in the decay phases of the $V$- and TESS-band light curves.  The HST stopped observing about 160~s into the fast UV rise of Flare Event 2 (Figure \ref{fig:lc_flare2}), after which the flaring TESS flux doubled before reaching a peak.  Presumably, the UV would have also continued to rise dramatically after the end of the HST observations.  TESS data at $\Delta t = 120$~s cadence are available for both flare events, but we do not show them here.

TESS \texttt{SAP\_FLUX}, $f$, is in data units.  We convert to $\Delta f / f_{\rm{preflare}}$ where $\Delta f = f(t) - f_{\rm{preflare}}$.  Then we convert to energy as follows.  The HST/FOS quiescent spectrum \citep{Augereau2006, Tristan2023} of AU Mic is a representative template of a M1e star, which is scaled to the quiescent $V$-band magnitude of CR Dra.   \citet{NLDS}\footnote{\url{https://www.stsci.edu/~inr/cmd.html}} and \citet{PMSU1} report $V=9.97$ for CR Dra, but the SIMBAD value is significantly brighter \citep[$V=9.46$;][]{Ducati2002}.    The yearly-averaged $V$ and $I_c$ magnitudes are derived from the Kamogata/Kiso/Kyoto Wide-field Survey (KWS) data (retrieved from the KWS website at \url{http://kws.cetus-net.org/~maehara/VSdata.py}).  The data points with large photometry errors ($> 0.08$ mag for $V$ band; $>0.05$ mag for $I_c$ band) are removed, and the data within each observing season are averaged.  At the time of the 2024 observations of CR Dra, the Johnson $V$-band magnitude is $\approx 10.1$ \citep[the KWS $V$ and $I_c$ band data were calibrated using $V$ and $I_c$ magnitudes of nearby stars in the Hipparcos main catalog;][]{Perryman1997}.  We thus scale to a $V$-band filter-weighted flux density \citep{Sirianni2005} corresponding to $V=10.1$ using a Johnson $V$-band zeropoint \citep{Willmer2018} and the photon-counting sensitivity curve from \citet{MA2006}.  Following \citet{Davenport2012}, we extend the HST/FOS spectrum to the TESS bandpass by scaling a dM1e optical template \citep{Bochanski2007} from the Sloan Digital Sky Survey and a spectrum of the M1 star HD 42581 from the NASA Infrared Telescope Facility catalog of cool star templates \citep{Cushing2005, Rayner2009}.
The photon-counting sensitivity curve of the TESS bandpass gives the quiescent flux of CR Dra to be $6.1 \times 10^{-13}$ erg cm$^{-2}$ s$^{-1}$ \AA$^{-1}$.  The flare equivalent durations \citep{Gershberg1972}, the distance \citep[$d = 20.3$ pc;][]{GaiaDR2} to CR Dra, and the TESS bandpass FWHM ($3982$ \AA) are used to calculate flare energies, $E_{\rm{TESS}}$, and peak luminosities.  An approximate transformation to energies calculated for a $T=9000$ K blackbody flare, as often done in the literature, is to multiply our bandpass energies by a factor of $\approx 5$; for a $T=10,000$ K blackbody, the factor is $\approx 6$.  The flare-only flux of Flare Event 2 (Figure \ref{fig:summaryfig}) is calculated by multiplying the value of $\Delta f / f_{\rm{preflare}} = 0.09$ by the TESS bandpass quiescent flux above.  

Several calculated quantities from the TESS light curves are summarized in Table \ref{table:TESS}.

\begin{deluxetable}{lccccccc}
\tabletypesize{\scriptsize}
\tablewidth{0pt} 
\tablecaption{Calculated Quantities from TESS ($\Delta t = 20$~s) Light Curves}\label{table:TESS}
\tablehead{
\colhead{Flare} & \colhead{$f_{\rm{preflare}}$}& \colhead{Peak $\Delta f / f_{\rm{preflare}}$} & \colhead{$t_{1/2}$} & \colhead{Impulsiveness$^{b}$} & \colhead{Equivalent Duration} & \colhead{Total Duration} & \colhead{$e^{-1}$ Duration} \\
\colhead{} & \colhead{TESS DN} & \colhead{ } & \colhead{s} & \colhead{$10^{-5}$ s$^{-1}$} & \colhead{s} & \colhead{s} & \colhead{s} \\ }
\startdata 
Flare Event 1 & 84354 & 0.051 & 911$^{a}$ & $4.5-5.6$ & 62 & 5500 & 880 \\
Flare Event 2 & 89163 & 0.734 & 650       & $113$ &     757 & 17,000 & 720 \\
\enddata
\tablecomments{$t_{1/2}$ is the FWHM of the light curve, and $e^{-1}$ duration is the elapsed time between the peak flux time and the time when the flux reaches a value of peak flux $ \div\ e$ \citep{Maehara2015, Namekata2017}. 
 $^{a}$Including both peaks in Flare Event 1 gives a full-width at 0.45 of the maximum of 1122~s.  $^b$Impulsiveness index calculated as peak $I_f$ divided by $t_{1/2}$. The total durations are roughly determined according to the time when the gradually decaying flux returns to the level of the pre-flare flux, $f_{\rm{preflare}}$, within the scatter. }
\end{deluxetable}

\section{Supplementary Spectra} \label{sec:supplementary_spectra}

\begin{figure}[h!]
\centering
\includegraphics[width=0.85\textwidth]{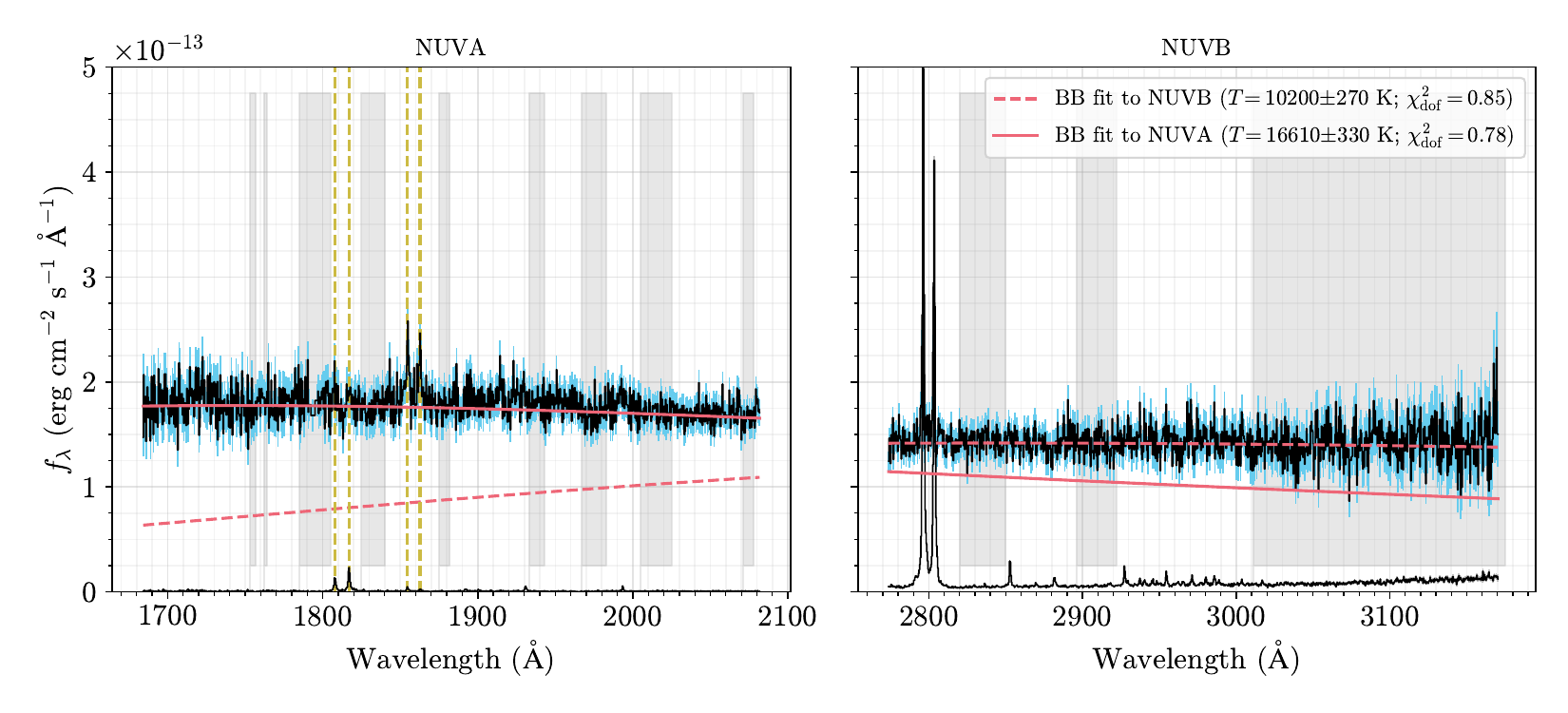}
\includegraphics[width=0.85\textwidth]{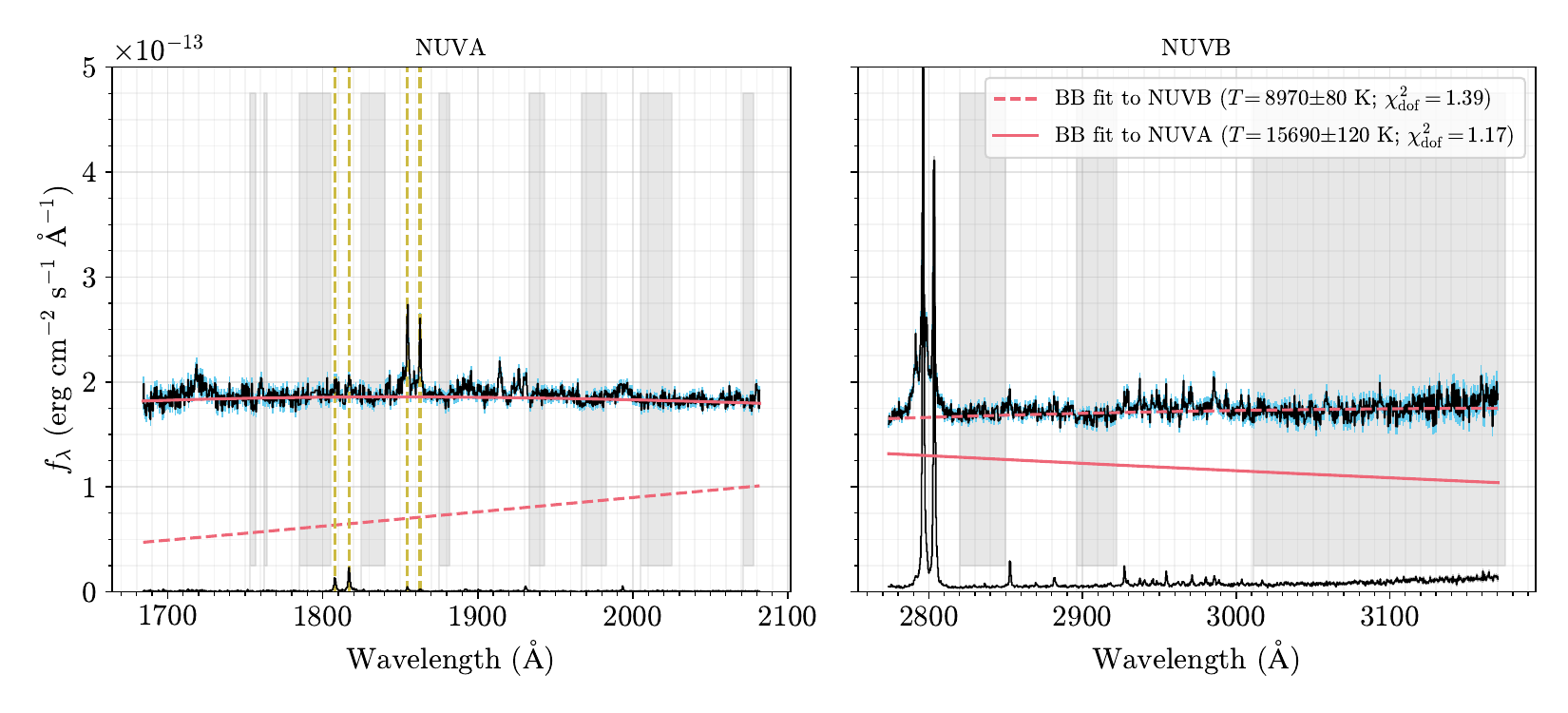}
\includegraphics[width=0.85\textwidth]{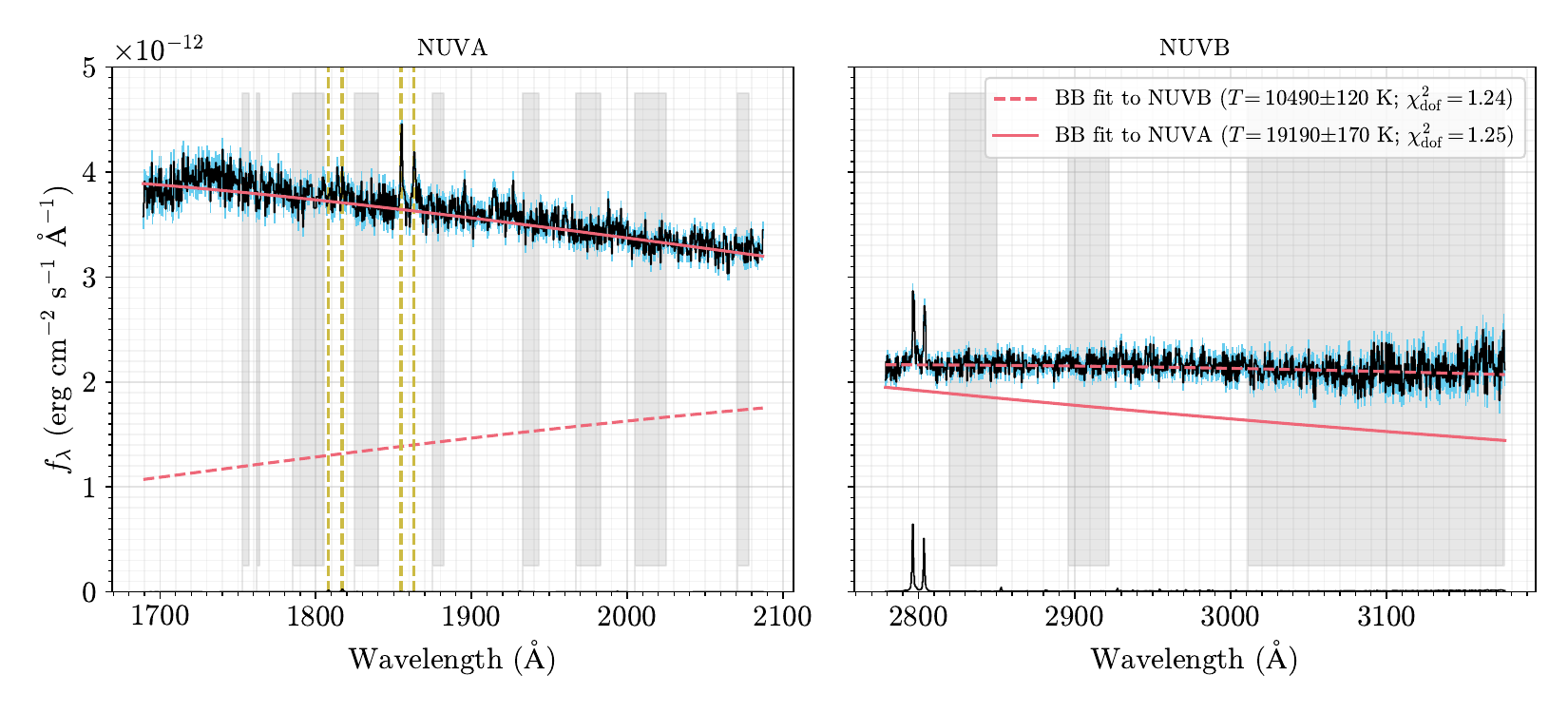}
\caption{ Flare spectra and blackbody fits (see text) from the first fast rise phase (UTC 2024-03-17 13:26:38.40 to 13:27:43.40) of Flare Event 1 (top panel) and over the entire impulsive phase  (UTC 2024-03-17 13:26:38.40 - 13:35:23.40) of Flare Event 1 (middle panel), and a shorter observation time (2024-05-17 11:46:43.75 - 11:47:13.75) just before  HST stopped observing Flare Event 2 (bottom panel).
 }\label{fig:other_bb}
\end{figure}

\begin{figure}[h!]
\centering
\includegraphics[width=0.85\textwidth]{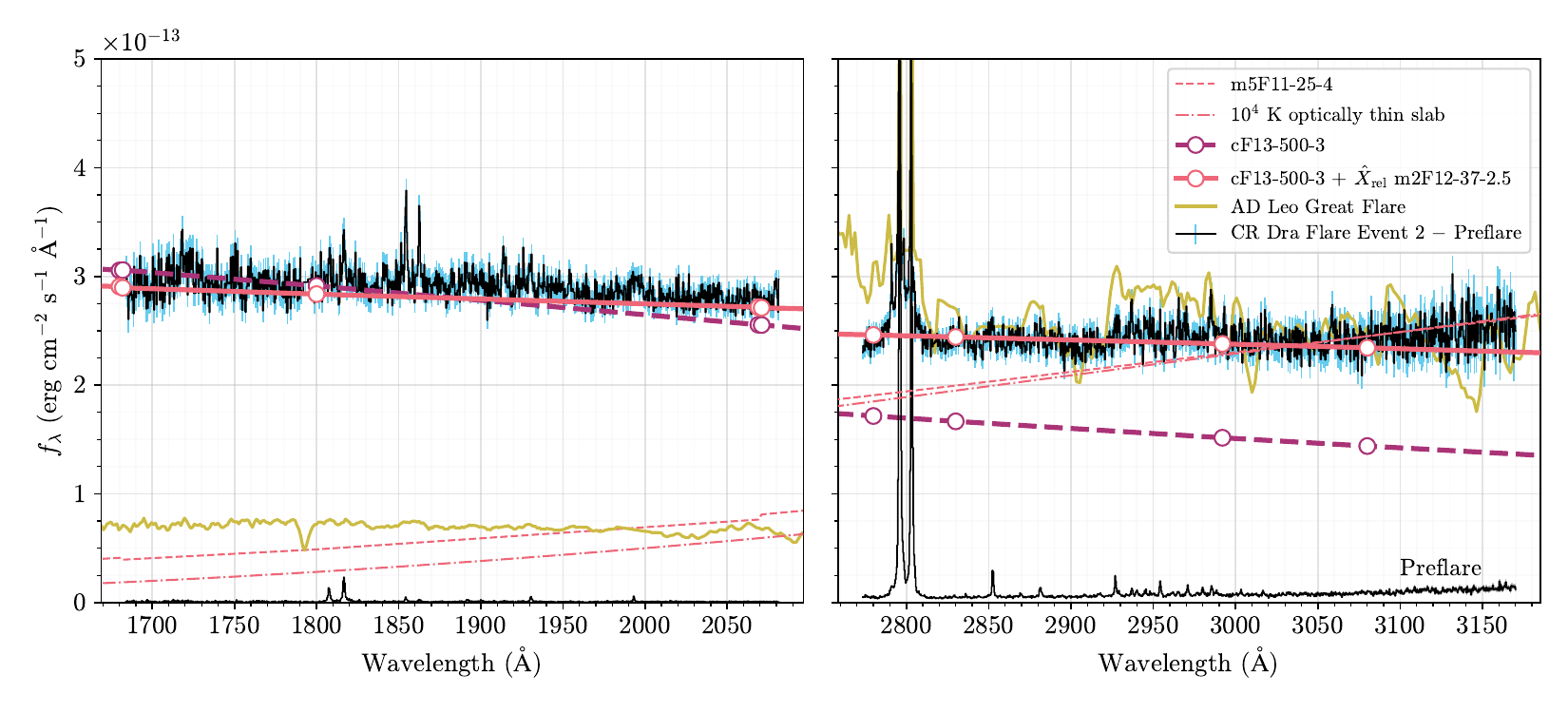}
\includegraphics[width=0.85\textwidth]{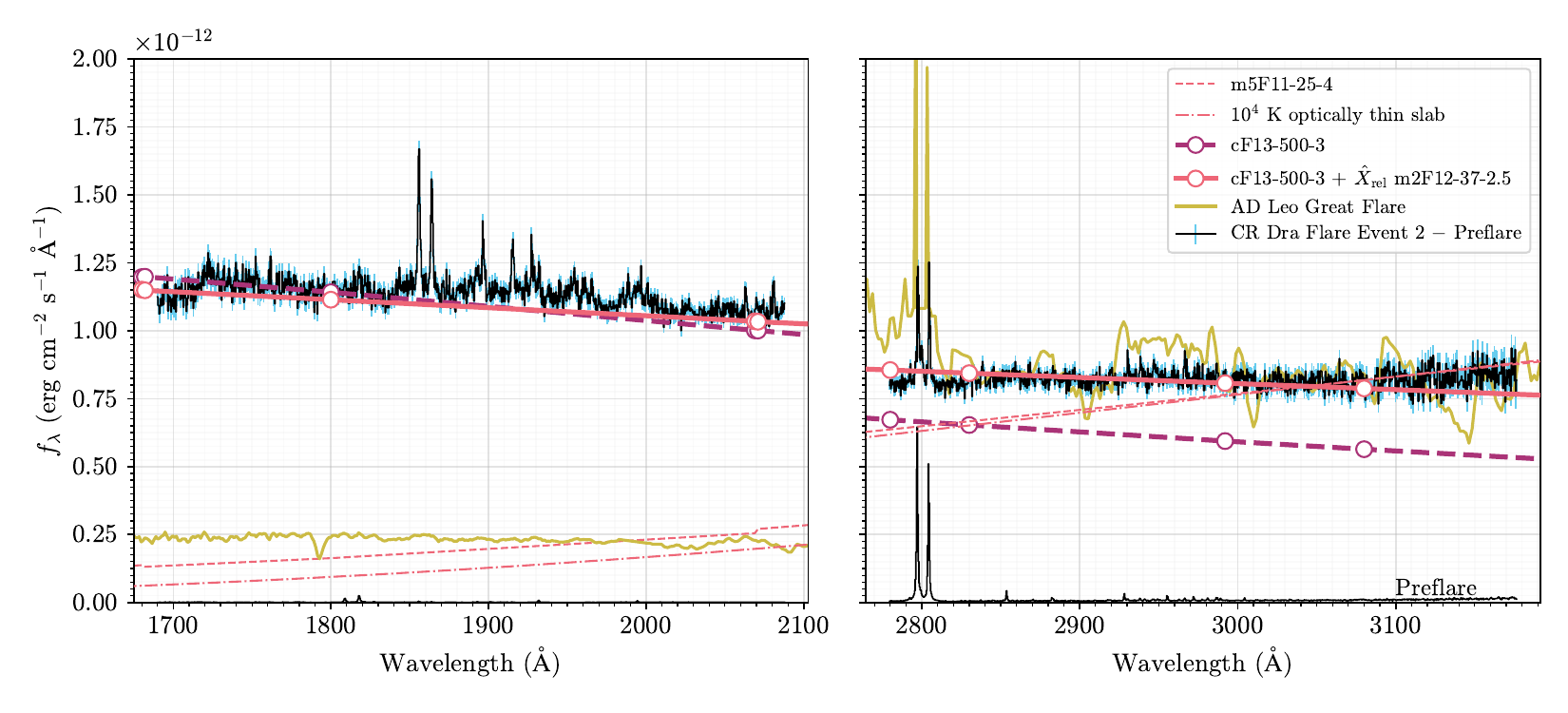}
\includegraphics[width=0.85\textwidth]{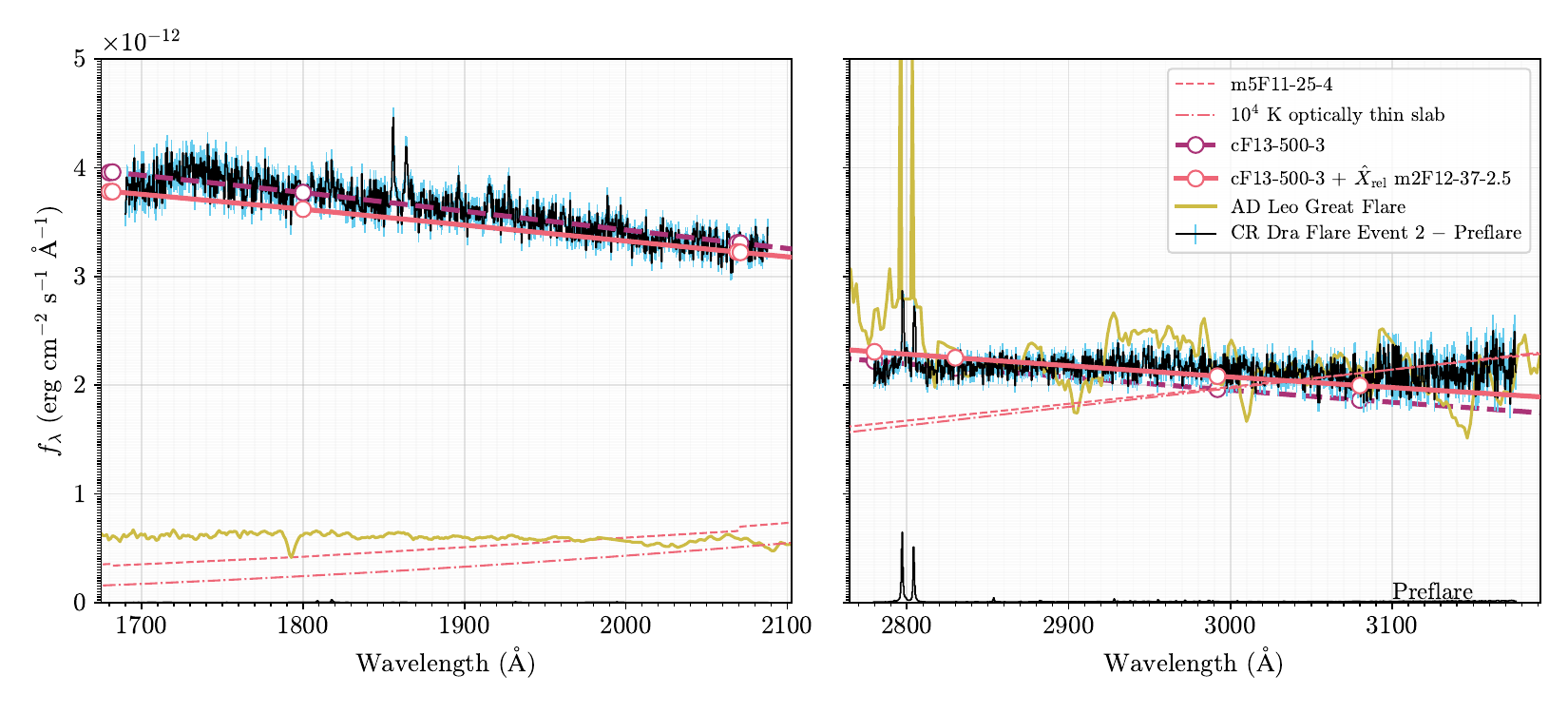}
\caption{ Same as Figure \ref{fig:RHD} except for times (UTC 2024-03-17 13:31:53.40 - 13:35:23.40) corresponding to the second major peak in Flare Event 1 (top panel),  an extraction of 120~s corresponding to the last six TESS short-cadence observations (2024-05-17 11:45:00.85 - 11:47:00.85) before HST stopped observing Flare Event 2 (middle panel), and a shorter observation time (2024-05-17 11:46:43.75 - 11:47:13.75) just before  HST stopped observing Flare Event 2 (bottom panel).
 }\label{fig:other_RHD}
\end{figure}

This appendix supplements Figures \ref{fig:spectra} and \ref{fig:RHD} with other flare spectra and their corresponding blackbody and RHD model fits at several interesting times during Flare Event 1 and 2.  These demonstrate how the best-fit blackbody temperatures and RHD filling factors differ compared to the integrated spectra over the second major peak in Flare Event 1 and over the entire fast rise phase interval that was observed in Flare Event 2.  For example, the spectrum from the last 30~s before HST stopped in Flare Event 2 is shown in the bottom panels of Figure \ref{fig:other_bb} and Figure \ref{fig:other_RHD}.  The blackbody fit to NUVA has a color temperature  of nearly $\hat{T}_{\rm{NUVA}} \approx 19,000$ K, and the value of $\hat{X}_{\rm{rel}}$ is $\approx 1.2$.  The satisfactory fits of the two component RHD model in the second major peak of Flare Event 1 (Figure \ref{fig:other_RHD} (top)), when the NUVA flux  was relatively fainter than in Flare Event 2, demonstrates the versatility of this approach.  The middle panel of Figure \ref{fig:other_RHD} demonstrates the fitting over the time interval corresponding exactly to the last 6 data points of TESS before HST stopped observations ($\hat{X}_{\rm{rel}} = 4.5$; Figure \ref{fig:summaryfig}).   The spectrum that is summed over the impulsive phase of Flare Event 1, consisting of two major peaks and their respective fast decay phases, is shown in Figure \ref{fig:other_bb} (middle).  This flare spectrum has the highest signal-to-noise ratios, but the relative fluxes in NUVA and NUVB vary considerably over the course of the impulsive phase (Figure \ref{fig:lcs}).  The flare spectrum over the first fast rise phase of Flare Event 1 (Figure \ref{fig:other_bb}) does not have any flux contributing from a previously decaying peak;  the NUVB is remarkably flat, and the NUVA has a rather large blackbody color temperature of $\hat{T}_{\rm{NUVA}} \approx 16,600$ K.

\software{astropy \citep{astropy:2022}
          }

\bibliography{biblio}{}
\bibliographystyle{aasjournal}

\end{document}